\documentclass[sigconf]{acmart}

\AtBeginDocument{%
  }

\copyrightyear{2025}
\acmYear{2025}
\setcopyright{cc}
\setcctype{by-nc}
\acmConference[UIST '25]{The 38th Annual ACM Symposium on User Interface Software and Technology}{September 28-October 1, 2025}{Busan, Republic of Korea}
\acmBooktitle{The 38th Annual ACM Symposium on User Interface Software and Technology (UIST '25), September 28-October 1, 2025, Busan, Republic of Korea}\acmDOI{10.1145/3746059.3747743}
\acmISBN{979-8-4007-2037-6/2025/09}

\usepackage{enumitem}
\usepackage{xspace}
\usepackage{pifont}

\usepackage{tabularx}
\usepackage{multirow}
\usepackage{dcolumn} 
\newcolumntype{d}[1]{D{.}{.}{#1}}
\usepackage{color, colortbl}

\definecolor{Gray}{gray}{0.95}
\definecolor{darkgreen}{rgb}{0, 0.5, 0}
\definecolor{darkred}{rgb}{0.8, 0, 0}

\usepackage{dblfloatfix}

\begin{document}

\title{EclipseTouch: Touch Segmentation on Ad Hoc Surfaces using Worn Infrared Shadow Casting}
\newcommand{\systemname}{EclipseTouch\xspace}

\author{Vimal Mollyn}
\authornote{Both authors contributed equally.}
\affiliation{
 \institution{Carnegie Mellon University}
 \city{Pittsburgh}
 \state{PA}
 \country{USA}}
\email{vmollyn@cs.cmu.edu}

\author{Nathan DeVrio}
\authornotemark[1]
\affiliation{
 \institution{Carnegie Mellon University}
 \city{Pittsburgh}
 \state{PA}
 \country{USA}}
\email{ndevrio@cmu.edu}

\author{Chris Harrison}
\affiliation{
 \institution{Carnegie Mellon University}
 \city{Pittsburgh}
 \state{PA}
 \country{USA}}
\email{chris.harrison@cs.cmu.edu}

\renewcommand{\shortauthors}{Mollyn, DeVrio \& Harrison}

\begin{abstract}
The ability to detect touch events on uninstrumented, everyday surfaces has been a long-standing goal for mixed reality systems. Prior work has shown that virtual interfaces bound to physical surfaces offer performance and ergonomic benefits over tapping at interfaces floating in the air. A wide variety of approaches have been previously developed, to which we contribute a new headset-integrated technique called \systemname. We use a combination of a computer-triggered camera and one or more infrared emitters to create structured shadows, from which we can accurately estimate hover distance (mean error of 6.9~mm) and touch contact (98.0\% accuracy). We discuss how our technique works across a range of conditions, including surface material, interaction orientation, and environmental lighting. 
\end{abstract}


\begin{teaserfigure}
  \includegraphics[width=\textwidth]{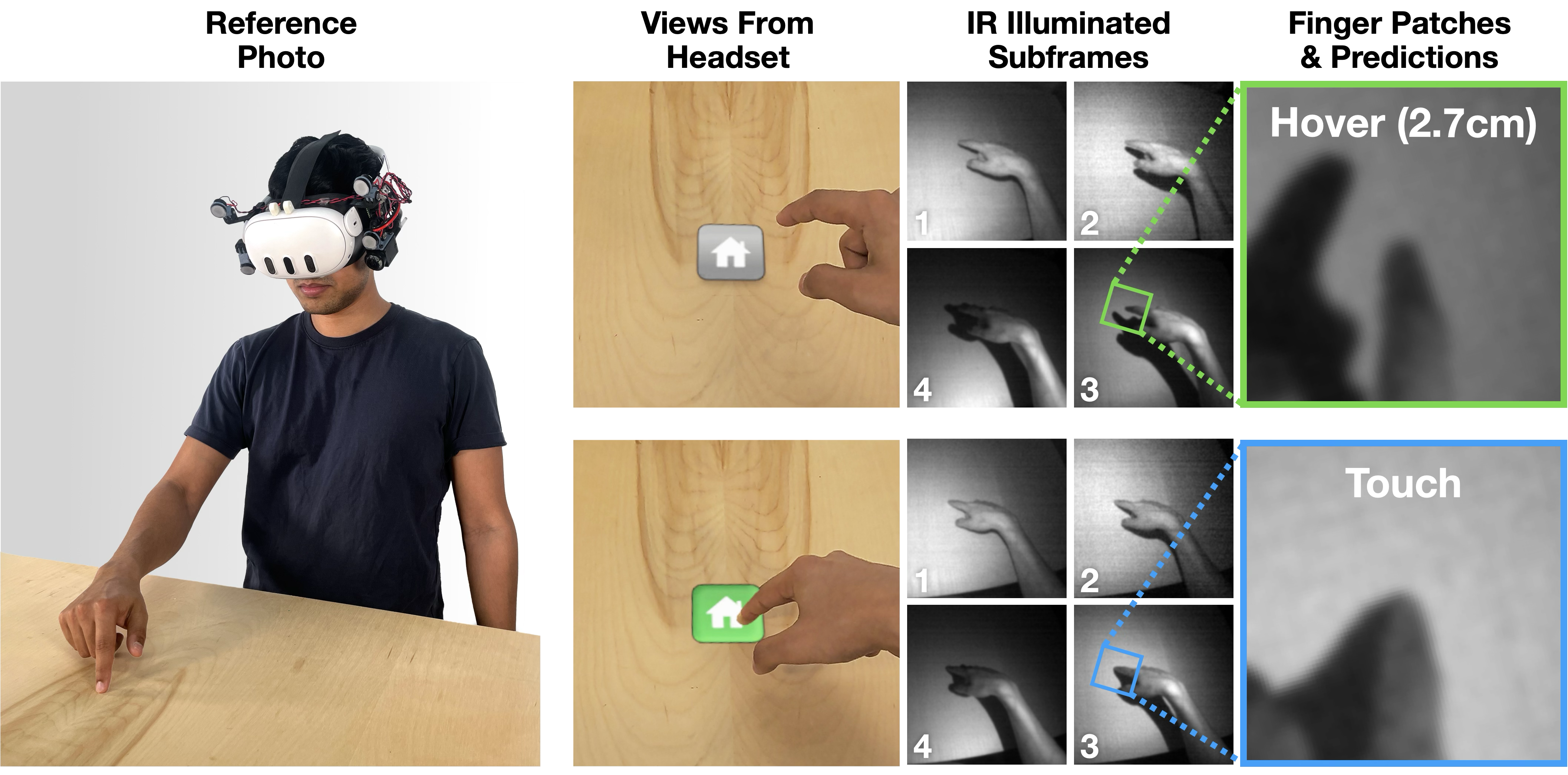}
  \caption{\systemname is a headset-integrated sensing approach for touch input on ad hoc surfaces. The headset illuminators create structured shadows in infrared (1/2/3/4), which our system uses to estimate touch contact and hover distance.}
  \label{fig:teaser}
  \vspace{2.5mm}
\end{teaserfigure}

\maketitle

\section{Introduction}
\begin{table*}[t]
    \centering
    \resizebox{\textwidth}{!}{
    \begin{tabular}{l l c c c c c c c c c c}
    \toprule
    \textbf{\begin{tabular}[l]{@{}l@{}}System \\ Name \end{tabular}} & \textbf{\begin{tabular}[l]{@{}l@{}}Sensor \\ Hardware \end{tabular}} & \textbf{\begin{tabular}[c]{@{}c@{}}Mobile vs. \\ Stationary\end{tabular}} & \textbf{\begin{tabular}[c]{@{}c@{}}Instrumented \\ Hands / Arms \end{tabular}} & \textbf{\begin{tabular}[c]{@{}c@{}}Supports \\ Multitouch\end{tabular}} & \textbf{\begin{tabular}[c]{@{}c@{}}World \\ Input\end{tabular}} & \textbf{\begin{tabular}[c]{@{}c@{}}Skin \\ Input\end{tabular}} & \textbf{\begin{tabular}[c]{@{}c@{}}Demonstrated \\ in Darkness\end{tabular}} & \textbf{\begin{tabular}[c]{@{}c@{}}Estimates \\ Hover Distance\end{tabular}} & \textbf{\begin{tabular}[c]{@{}c@{}}Touch \\ Accuracy\end{tabular}} \\
    
    \midrule
    
    TapLight \cite{streli_structured_2023}              & IR Camera, Structured Light           & \cellcolor[HTML]{D9EAD3}Mobile            & \cellcolor[HTML]{D9EAD3}No        & \cellcolor[HTML]{F4CCCC}No        & \cellcolor[HTML]{D9EAD3}Yes        & \cellcolor[HTML]{F4CCCC}No        & \cellcolor[HTML]{F4CCCC}No        & \cellcolor[HTML]{F4CCCC}No        & 95.3\% \\
    OmniTouch \cite{harrison_omnitouch_2011}            & Depth Camera                          & \cellcolor[HTML]{D9EAD3}Mobile            & \cellcolor[HTML]{D9EAD3}No        & \cellcolor[HTML]{D9EAD3}Yes       & \cellcolor[HTML]{D9EAD3}Yes       & \cellcolor[HTML]{D9EAD3}Yes        & \cellcolor[HTML]{F4CCCC}No        & \cellcolor[HTML]{F4CCCC}No        & 96.5\% \\
    EgoPressure \cite{zhao_egopressure_2024}            & RGB Camera                            & \cellcolor[HTML]{D9EAD3}Mobile            & \cellcolor[HTML]{D9EAD3}No        & \cellcolor[HTML]{D9EAD3}Yes       & \cellcolor[HTML]{D9EAD3}Yes        & \cellcolor[HTML]{F4CCCC}No        & \cellcolor[HTML]{F4CCCC}No        & \cellcolor[HTML]{F4CCCC}No        & n.r. \\
    MRTouch \cite{xiao_mrtouch_2018}                    & Depth \& IR Camera                    & \cellcolor[HTML]{D9EAD3}Mobile            & \cellcolor[HTML]{D9EAD3}No        & \cellcolor[HTML]{F4CCCC}No        & \cellcolor[HTML]{D9EAD3}Yes        & \cellcolor[HTML]{F4CCCC}No       & \cellcolor[HTML]{F4CCCC}No        & \cellcolor[HTML]{F4CCCC}No        & 96.5\% \\
    PressureVision++ \cite{grady_pressurevision_2024}   & RGB Camera                            & \cellcolor[HTML]{F4CCCC}Stationary        & \cellcolor[HTML]{D9EAD3}No        & \cellcolor[HTML]{D9EAD3}Yes       & \cellcolor[HTML]{D9EAD3}Yes        & \cellcolor[HTML]{F4CCCC}No        & \cellcolor[HTML]{F4CCCC}No        & \cellcolor[HTML]{F4CCCC}No        & 89.3\% \\
    PlayAnywhere \cite{wilson_playanywhere_2005}        & IR Camera, IR LED                     & \cellcolor[HTML]{F4CCCC}Stationary        & \cellcolor[HTML]{D9EAD3}No        & \cellcolor[HTML]{F4CCCC}No        & \cellcolor[HTML]{D9EAD3}Yes        & \cellcolor[HTML]{F4CCCC}No        & \cellcolor[HTML]{F4CCCC}No        & \cellcolor[HTML]{F4CCCC}No        & n.r. \\
    Matsubara et al. \cite{matsubara_touch_2017}        & IR Camera, IR LEDs                    & \cellcolor[HTML]{F4CCCC}Stationary        & \cellcolor[HTML]{D9EAD3}No        & \cellcolor[HTML]{F4CCCC}No        & \cellcolor[HTML]{D9EAD3}Yes        & \cellcolor[HTML]{F4CCCC}No        & \cellcolor[HTML]{F4CCCC}No        & \cellcolor[HTML]{F4CCCC}No        & 96.1\% \\
    EgoTouch \cite{mollyn_egotouch_2024}                & RGB Camera                                  & \cellcolor[HTML]{D9EAD3}Mobile            & \cellcolor[HTML]{D9EAD3}No        & \cellcolor[HTML]{D9EAD3}Yes       & \cellcolor[HTML]{F4CCCC}No       & \cellcolor[HTML]{D9EAD3}Yes        & \cellcolor[HTML]{F4CCCC}No        & \cellcolor[HTML]{F4CCCC}No        & 94.9\% \\
    Shadow Touch \cite{liang_shadowtouch_2023}          & RGB Camera, White LED                 & \cellcolor[HTML]{D9EAD3}Mobile            & \cellcolor[HTML]{F4CCCC}Yes       & \cellcolor[HTML]{D9EAD3}Yes       & \cellcolor[HTML]{D9EAD3}Yes        & \cellcolor[HTML]{F4CCCC}No        & \cellcolor[HTML]{F4CCCC}No        & \cellcolor[HTML]{F4CCCC}No        & 99.1\% \\
    \textbf{EclipseTouch (Ours)} & \textbf{IR Camera, IR LEDs} & \textbf{\cellcolor[HTML]{D9EAD3}Mobile} & \textbf{\cellcolor[HTML]{D9EAD3}No} & \textbf{\cellcolor[HTML]{D9EAD3}Yes} & \textbf{\cellcolor[HTML]{D9EAD3}Yes} & \textbf{\cellcolor[HTML]{D9EAD3}Yes} & \textbf{\cellcolor[HTML]{D9EAD3}Yes} & \textbf{\cellcolor[HTML]{D9EAD3}Yes (6.9 mm error)} & \textbf{98.0\%} \\
    
    \toprule
    \end{tabular}}
    \vspace*{1mm}\caption{Overview of key related work. Green is a positive attribute; red is negative. Refer to individual papers for study details.} 
    \label{tab:touch-systems-overview}
    \vspace*{-5mm}
\end{table*}

Mixed reality (XR/AR) headsets are becoming more widespread, with increasing consumer excitement about forthcoming glasses-like form factors. However, navigating user interfaces on these devices generally requires users either to carry accessory controllers everywhere they go, or be limited to poking and swiping at interfaces in the air.

\sloppy
A complementary, long-envisioned interaction option is to ground virtual interfaces to real-world surfaces. One of the most reliable ways to achieve such interactions is to instrument surfaces with sensors. While robust, it is not feasible to do this for every surface in the world. In a future with highly mobile wearers of AR glasses, users may wish to temporarily and opportunistically appropriate surfaces for touch interaction. For this reason, there exists a significant body of work on ad hoc touch input without instrumentation of the environment. In most cases, this requires instrumenting the user instead, often with a special-purpose accessory device. Even when such sensors could be plausibly integrated into future smartwatches, it is most common for people to wear watches on their non-dominant hand.

We believe the ideal sensing method should integrate directly into the headset/glasses so that the user only needs to carry a single, self-contained device (i.e., the glasses). Only a handful of methods, which we discuss in Related Work, achieve this property. To this, we add two additional practical constraints: robust operation 1) across a wide variety of materials, including the user's skin; and 2) across environmental conditions (dark, bright, noisy, moving, etc.).

In this work, we present \systemname, a new headset-integrated technique for ad hoc touch sensing. Our system leverages the well-known phenomena of shadow casting, utilized in prior touch sensing work, but never demonstrated in a single worn device. More specifically, \systemname uses an infrared, egocentric headset camera, which captures shadows cast by one or more synchronized infrared illuminators on the headset (Figure~\ref{fig:pipelineandhardware}). As the geometry between the camera and illuminators are fixed, these shadows inherently capture a finger's distance from a surface, including direct contact (Figure~\ref{fig:distanceestimationexample}). Importantly, our approach must first filter out shadows cast by extraneous light sources in order to be robust. At the core of our system is an optimized deep neural network that has an inference time of 0.47~ms on an Apple M2 processor (used in the Apple Vision Pro). The result is a method that works "out of the box" and requires no pre-registration or calibration of the environment, surface, or user. Our approach works across a wide array of common surfaces, as well as lighting conditions, from bright to pitch-black. \systemname can also be readily integrated into several popular XR headsets that already contain the requisite sensing hardware and compute. Taken together, this set of capabilities sets it apart from prior work, even those relying on similar phenomena.

\section{Related Work}\label{sec:related work}

In this section, we review prior systems that have used techniques relevant to \systemname. We begin with systems that examined the problem of ad hoc surface touch detection. For these, we start with instrumented environments (the systems least similar to our approach), progress to arm-worn mobile systems, and finally to mobile systems that do not require arm instrumentation (most similar to our approach). We conclude this section with a review of systems that specifically used shadows for touch tracking.

We emphasize that although active-illumination shadow tracking has been used previously --- including for the exact same use case as this present work --- our particular instantiation offers a favorable mix of capabilities not demonstrated in any prior work (Table \ref{tab:touch-systems-overview}), including: 

\begin{itemize}[left=0pt,label={\scriptsize$\bullet$}]
    \item Uses hardware already present in modern headsets (cameras and illuminators). 
    \item Requires no instrumentation of the user's arms (i.e., bare hands).
    \item Enables ad hoc input for commonplace surfaces, including the user's skin.
    \item Works across lighting conditions, including in complete darkness.
    \item Works "out of the box", requiring no pre-registration or calibration of the surface, user, or environment.
    \item Offers high touch input accuracy (98.0\% touch segmentation, 6.9~mm hover distance estimation mean error). 
\end{itemize}

\subsection{Detecting Touch with Instrumented Environments}

The most straightforward way to add touch tracking to a surface is to instrument that surface directly. This is a well-explored area of research that we will not cover in depth for brevity. Some of the most common techniques have included placing microphones \cite{pham_tangible_2002, ono_touch_2013,scratchInput} or LIDAR \cite{strickon_tracking_1998,SurfaceSight} on the surface, or cameras above the surface. 
For cameras, there have been many different optical approaches. For example, thermal cameras have been used to detect changes in heat left behind on the surface after touch \cite{iwai_heat_2005, kurz_thermal_2014, larson_heatwave_2011, funk_interactive_2015}. RGB cameras have been used to capture images of the fingernail and its color change during presses \cite{marshall_pressing_2009, sugita_touch_2008, agarwal_high_2007, chen_estimating_2020} or even estimate fingertip pressure \cite{chen_estimating_2020, grady_pressurevision_2024, grady_pressurevision_2022}. Perhaps the most popular technique has been to use fixed depth cameras operating above surfaces. After Benko and Wilson's works that pioneered this approach \cite{benko_depthtouch_2009, wilson_using_2010}, many other iterations have improved performance using flood-fill algorithms \cite{xiao_desktop_2017}, combining infrared camera data \cite{xiao_direct_2016}, and applying machine learning \cite{fan_reducing_2022}. Other lesser used optical approaches include multi-path interference from infrared depth cameras \cite{shen_farout_2021, xia_halotouch_2025}, laser speckle imaging \cite{pei_forcesight_2022}, and imaging reflections of the finger in mirrors and glossy surfaces \cite{yoo_symmetrisense_2016, pak-kiu_chung_mirrortrack_2008}. 
A major drawback of all these systems is that they need to instrument every potential surface or environment that the user may wish to interact with, which scales poorly. For this reason, research has looked into instrumenting the user with sensors to detect touches on ad-hoc surfaces, which we review next.

\subsection{Detecting Touch with Finger/Hand/Arm- Mounted Sensors}

When a user touches a surface, their touching finger produces a number of characteristic signals that could be used to detect touch contact. For instance, IMUs attached to the fingernail can detect spikes in deceleration that occur when a finger touches a surface \cite{oh_anywheretouch_2017, oh_fingertouch_2020, shi_ready_2020}, but are cumbersome for users to wear and recharge. More popular is an IMU ring form factor, which could be used to detect touch \cite{gu_accurate_2019, gu_qwertyring_2020} and mouse-like 2D inputs \cite{liang_dualring_2021, shen_mousering_2024, kienzle_lightring_2014}. Beyond rings, researchers have also explored using IMUs placed on the wrist (like a smartwatch) to detect tap events \cite{meier_tapid_2021}, however this signal is less robust.

Fingers also passively produce characteristic acoustic signals when tapping or swiping on a surface, which travel through the body and can be sensed with sensors mounted on the arm \cite{harrison_skinput_2010, masson_whichfingers_2017, gong_acustico_2020, kim_soundscroll_2024}. Alternatively, touches can be sensed through changes in reflections of actively emitted acoustic signals. For instance, Mujibiya et al. \cite{mujibiya_sound_2013}, SoundTrak \cite{zhang_soundtrak_2017}, and VersaTouch \cite{shi_versatouch_2020} emitted ultrasonic waves through transducers placed on the arm and finger to detect contact and pressure on the skin. 

For detecting touches to the skin, one accurate approach that researchers have explored is detecting changes to RF signals transmitted through the user's body. AtaTouch \cite{atatouch} detected subtle finger pinches by sensing changes in impedance between an antenna and the user's body. SkinTrack \cite{zhang_skintrack_2016} used two devices, a signal-emitting ring and a wristband receiver, with the body as an electrical waveguide to sense finger touch on the skin. ActiTouch \cite{zhang_actitouch_2019} and ElectroRing \cite{kienzle_electroring_2021} work in the same way, but move the transmitter and receiver to more convenient form factors. Finally, Z-Ring \cite{waghmare_z-ring_2023} used a single electrode ring that acted as a transceiver to sense bio-impedance changes in the hand. While these approaches are accurate, they are inherently limited to operation on conductive objects, such as the human body (but not most walls or furniture).
A key limitation of all these prior systems is that they require instrumentation of the user's finger, hand or arm, in addition to the XR headset.

\begin{figure*}[t]
     \centering
     \includegraphics[width=0.9\linewidth]{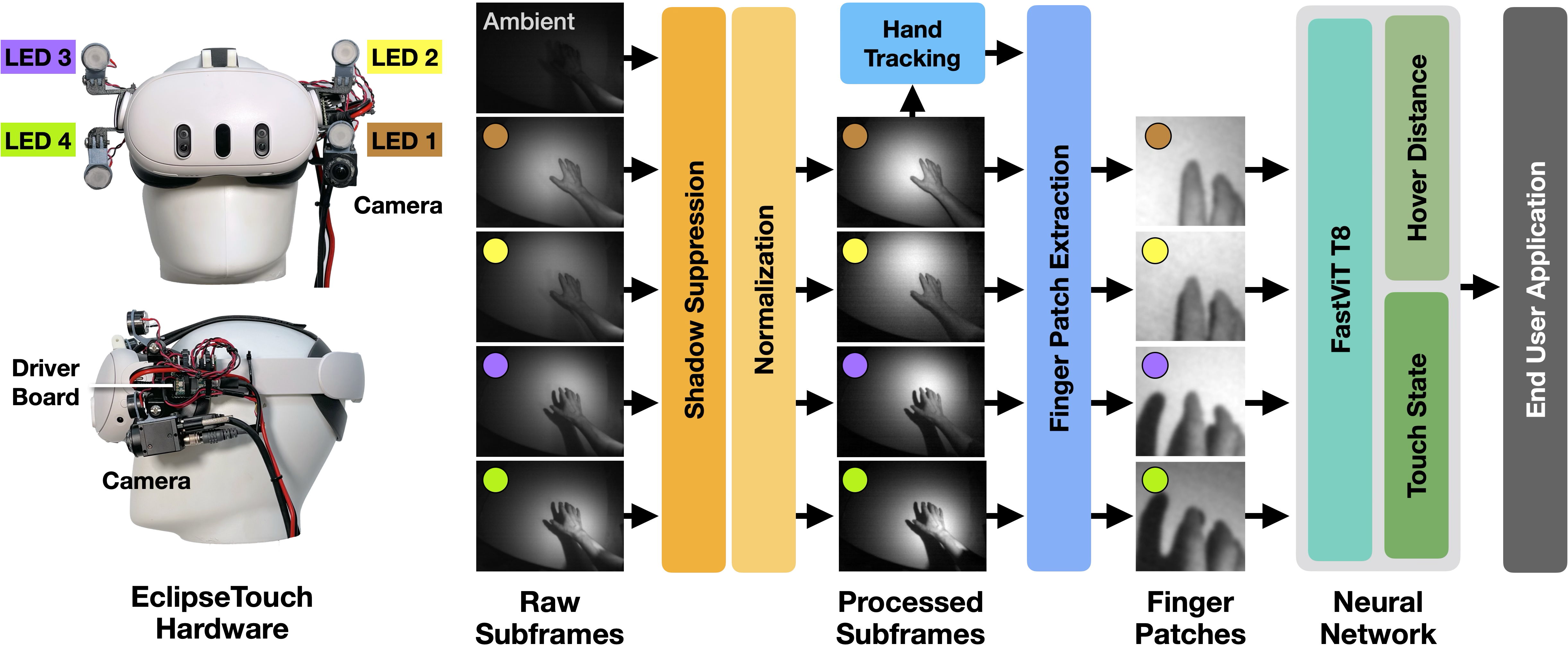}
     \caption{High-level overview of \systemname's experiment hardware and software pipeline. }
     \label{fig:pipelineandhardware}
\end{figure*}
 
\subsection{Detecting Touch without Finger/Hand/Arm- Mounted Sensors}
Rather than instrument a user's hands with a device (likely special-purpose, as even smartwatches are rarely worn on the dominant hand), a more practical approach would be to have all necessary hardware contained within the XR headset/glasses. Dominant among these approaches is to use headset mounted cameras with computer vision to detect touches on surfaces. 

As most modern headsets come with built-in 3D hand and world mesh tracking support, research has looked into inferring surface touch using this data. However, the world mesh modern headsets build is not currently accurate enough for touch detection purposes. For this reason, TriPad \cite{dupre_tripad_2024} required users to calibrate and define surfaces; hand tracking and dwells were used to instantiate touch planes and touches were detected by measuring fingertip proximity to the created plane. Richardson et al. \cite{richardson_decoding_2020} and Streli et al. \cite{streli_touchinsight_2024} focused on surface typing and used neural networks to analyze patterns of hand motion to detect surface taps with high accuracy. However, both these systems required pre-registered surfaces to work, and could not detect stateful touches (including hover and touch-ups).

Another approach that has seen the most success in the past is using depth cameras. OmniTouch \cite{harrison_omnitouch_2011} and Imaginary Phone \cite{gustafson_imaginary_2011} were early among these efforts and used depth cameras to track the fingers and detect touches on the palm and other surfaces using a combination of flood-filling and contour detection. MRTouch \cite{xiao_mrtouch_2018} combined depth sensor data with infrared reflectivity data from a Microsoft HoloLens to improve touch detection accuracy. 
The problem with depth cameras, even today, is noisy signal --- the difference between a finger touching vs. slightly hovering above a surface is hard to distinguish. For this reason, OmniTouch required users to lift fingers 20 mm above the surface to reliably separate hovering from touch events. This is awkward, and not like how one scrolls or types on their touchscreen devices. Moreover, depth cameras have other drawbacks including higher power consumption, lower resolution and lower framerates. Taplight \cite{streli_structured_2023} estimated fingertip depth and surface contact by using remote vibrometry (laser speckle sensing) obtained from a headset mounted laser and monochrome camera. This approach, while accurate in some contexts, does not function on many common surface materials, and can additionally fail from user head motion and multiple inputting fingers. 

Most relevant to \systemname is recent work on detecting surface contact using only headset cameras. This is challenging, as the finger directly occludes the point of touch contact, and the system must adapt to surface materials, lighting conditions, touch types, and skin tones. PressureVision \cite{grady_pressurevision_2022} and later PressureVision++ \cite{grady_pressurevision_2024} described deep learning-based approaches to detecting contact pressure of the hand to a surface, albeit with a fixed desk-mounted camera and restricted lighting and surface conditions. EgoPressure \cite{zhao_egopressure_2024} extended this work to headset-mounted cameras by collecting a large egocentric hand pressure estimation dataset, but their system ran offline (not in real-time) and required ambient illumination. In our prior work EgoTouch \cite{mollyn_egotouch_2024}, we demonstrated a system for detecting on-skin touch and force using only headset RGB cameras, by looking at patterns of skin deformation, color change and shadow convergence. Similar to EgoTouch, PalmPad \cite{palmpadchi2025} could also detect touches to the skin with a headset-mounted RGB camera, however it could not estimate force and only supported the palm. Both EgoTouch and PalmPad ran in real-time, but only worked on the skin, required ambient illumination (i.e. did not work in the dark), and were susceptible to false positives from extraneous shadows in the environment.  

The fundamental problem with all these prior approaches is that they rely on existing sources of illumination. However, ambient light is uncontrolled --- it can be diffused, harsh, oblique, bright, multicolored, single/multi-point, and even non-existent (dark). With \systemname, we sought to overcome these limitations by controlling the shadows cast by the finger on the surface. 

\subsection{Detecting Touch using Shadows}

To conclude our literature review, we now specifically discuss systems that have leveraged shadows to track user interactions, as this most closely relates to our technical approach. 

Starting with seminal work, we have Myron Krueger's Videoplace \cite{krueger_videoplaceartificial_1985}, which utilized user silhouettes for interactivity (though not true shadows). As far as we are aware, the earliest known work to leverage "shadow shape analysis" at the fingers for \textit{touch input} is Andy Wilson's PlayAnywhere \cite{wilson_playanywhere_2005}. This system used a camera and illuminator that was fixed with respect to the input plane by virtue of the system being placed on a table. This system laid the conceptual groundwork for using shadows for detecting touch, measuring the decreasing distance between the shadow and fingertip. 

More recent touch systems employing fixed cameras and fixed illuminators include ShadowReaching \cite{shoemaker_shadow_2007}, Iacolina et al. \cite{iacolina_improving_2011}, and Thomas \cite{thomas_camera_2013}. Matsubara et al. \cite{matsubara_touch_2017} and Niikura et al. \cite{niikura_touch_2016} used the same system, which featured two fixed infrared illuminators and a fixed camera capturing image pairs (one for each active illuminator). More ambitious is to have only a fixed camera and rely on existing natural or artificial light sources (i.e., no special illuminators). Adajania et al. \cite{adajania_virtual_2010}, Paper Piano \cite{vishal_paper_2017},
ShadowSense \cite{hu_shadowsense_2020}, and 
Posner et al. \cite{posner_single_2012} use this approach. However, with no control of the positioning between the camera and light sources (one or many, harsh or diffuse, bright or dim, etc.), input tends to be brittle, working well in some cases and failing in others.

Most similar to \systemname, is ShadowTouch \cite{liang_shadowtouch_2023}. Unlike the above prior work, ShadowTouch is worn and mobile (i.e., not reliant on fixed external infrastructure). Like \systemname, the system used an egocentric headset camera. Unlike \systemname, ShadowTouch requires a special-purpose LED wristband. As there is no synchronization between the LED and camera, the LED is persistently lit, which is prohibitively energy consumptive for a wearable. Additionally, the authors note that they do "not have a special design to alleviate the ambient light interference" \cite{liang_shadowtouch_2023}, an important part of our pipeline. We also move beyond ShadowTouch in terms of evaluation generalizability. In ShadowTouch's study, only three light-colored and matte surfaces are tested, only in a horizontal setting, and in one typically-lit environment. In this work, we test 12 surface materials (Figure~\ref{fig:materialstested}), both dark and light, and matte and reflective. We also test vertical and horizontal orientations, and across three lighting conditions (typical lighting, bright, and dark). We note that our evaluation results show (Section \ref{sec:results}) both systems to be of comparable accuracy; in other words, we achieve the same accuracy without the need for a special wearable. 

\section{Implementation}
\begin{figure*}[t]
     \centering
     \includegraphics[width=\linewidth]{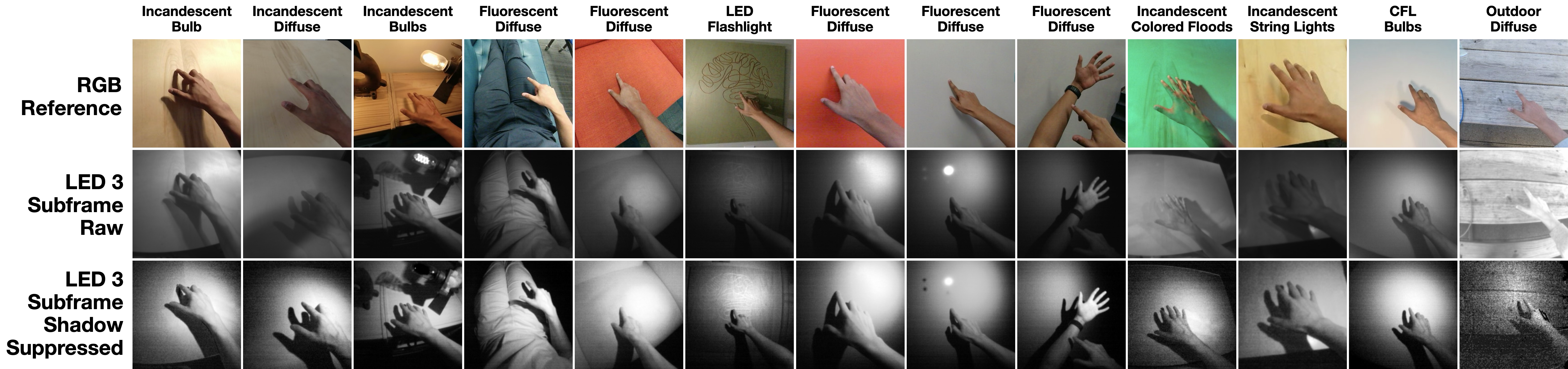}
     \caption{Examples of ambient shadow suppression output across a variety of materials, lighting types (incandescent, fluorescent tubes, CFL, LED, sun), and illumination conditions (diffuse, floodlight, point source). Note that shadows visible in visible light (RGB Reference) may be invisible in infrared and vice-versa. Figure \ref{fig:shadowsuppressionprocess} provides a step-by-step example of shadow suppression.}
     \label{fig:shadowsuppression}
\end{figure*}
As a prototype platform, we instrumented a Meta Quest 3. In the following subsections, we step through the various hardware and software components that make up \systemname. 

\subsection{Camera}
We affixed a HT-SUA33GM-T1V-C USB 3.0 camera \cite{noauthor_ht-sua33gm-t1v-c_nodate} to the bottom left of our prototype headset (Figure \ref{fig:pipelineandhardware}). This global shutter camera has a modest resolution of 640x480 pixels, but is sufficient for our needs (i.e., our shadows are large and textureless, not requiring high detail). We fitted the camera with an 850~nm bandpass filter, matching the wavelength of our LED illuminators (discussed next). Even with this filter, other light sources can cast visible shadows at 850~nm, most notably the sun, as well as incandescent and halogen lights. To account for this, our pipeline includes an extraneous shadow suppression process, described in Section~\ref{subsection:shadowsupression}. As we rely on only a narrow frequency band of light, we purposely selected a monochrome camera with a wideband CMOS sensor (with no Bayer pattern, which would preclude imaging infrared light). Finally, this camera features an external hardware trigger, which we use to synchronize with our LED illuminators. The camera streams video over USB to a laptop, where our software runs. The maximum framerate of the camera is 791 FPS, though we operate it about half this speed to increase exposure time. 

\subsection{Illumination Geometry}

As already discussed, \systemname relies on active illumination to create structured shadows. We use LEDs with a wavelength of 850~nm (infrared), that are safe and invisible to humans, even in darkness. The LEDs are rated at 3 W, but we drive them at 1.1~W as a power conservation measure.  Importantly, our prototype was built to serve as a vehicle for investigation. As such, we over-provisioned our headset with LEDs in each of the four corners. For the corner with the camera, the LED is placed directly above the camera. This arrangement can be seen in Figure \ref{fig:pipelineandhardware}. As we will evaluate and discuss later, some illuminator locations are more valuable than others, and so a commercial implementation would use fewer illuminators, potentially just one (as can be seen with our final prototype in Figure \ref{fig:new_prototype}).

\subsection{Driver Board}

We use a Teensy 3.2 microcontroller and custom MOSFET LED driver board to precisely control the timing of our LED illuminators and camera frames (microsecond precision). Our Teensy firmware uses the following five-step LED "firing sequence": No LEDs on, LED 1 on, LED 2 on, LED 3 on, and LED 4 on. Each step in the sequence has a duration of 2.5~ms, and at each step the camera is triggered using its external pinouts. The firing sequence loops continuously, producing a 400 FPS raw video stream. 

Our camera uses an exposure time of 2.4~ms. We note that this is a comparatively short exposure time for a camera with a small sensor (1/5.6"), but this is not an issue for \systemname because we are actively illuminating the scene. Furthermore, human skin is reflective in infrared, and the user's hands are never more than 70 cm away from the headset. As can be seen in Figures \ref{fig:shadowsuppression} and \ref{fig:materialstested}, the hand is readily seen, and the shadows cast are crisp and dark. 

\subsection{Video Stream}
On our laptop receiving the 400 FPS camera stream, we read five frames at a time (i.e., a complete firing sequence), and composite this data into a new, singular frame containing multiple illumination sources: five 640x480 images side-by-side in a row. Thus, this new composited stream has a framerate of 80 FPS. The time between the start of the first frame (no illumination) and the end of the last frame (LED 4 on) is approximately 12.5~ms. This short duration means that even when the hands are in motion, the image can be stacked for image processing as though they were taken at essentially the same moment in time. 

\subsection{Extraneous Shadow Suppression}\label{subsection:shadowsupression}
In addition to our infrared LEDs, other light sources with 850~nm wavelengths will cast finger shadows. Notable light sources include the sun, as well as some artificial lights, including incandescent and halogen bulbs (see example shadows in Figure~\ref{fig:shadowsuppression}). These shadows can generate false events, and so it is desirable to filter them out. 

\begin{figure}[b]
     \centering
     \includegraphics[width=\linewidth]{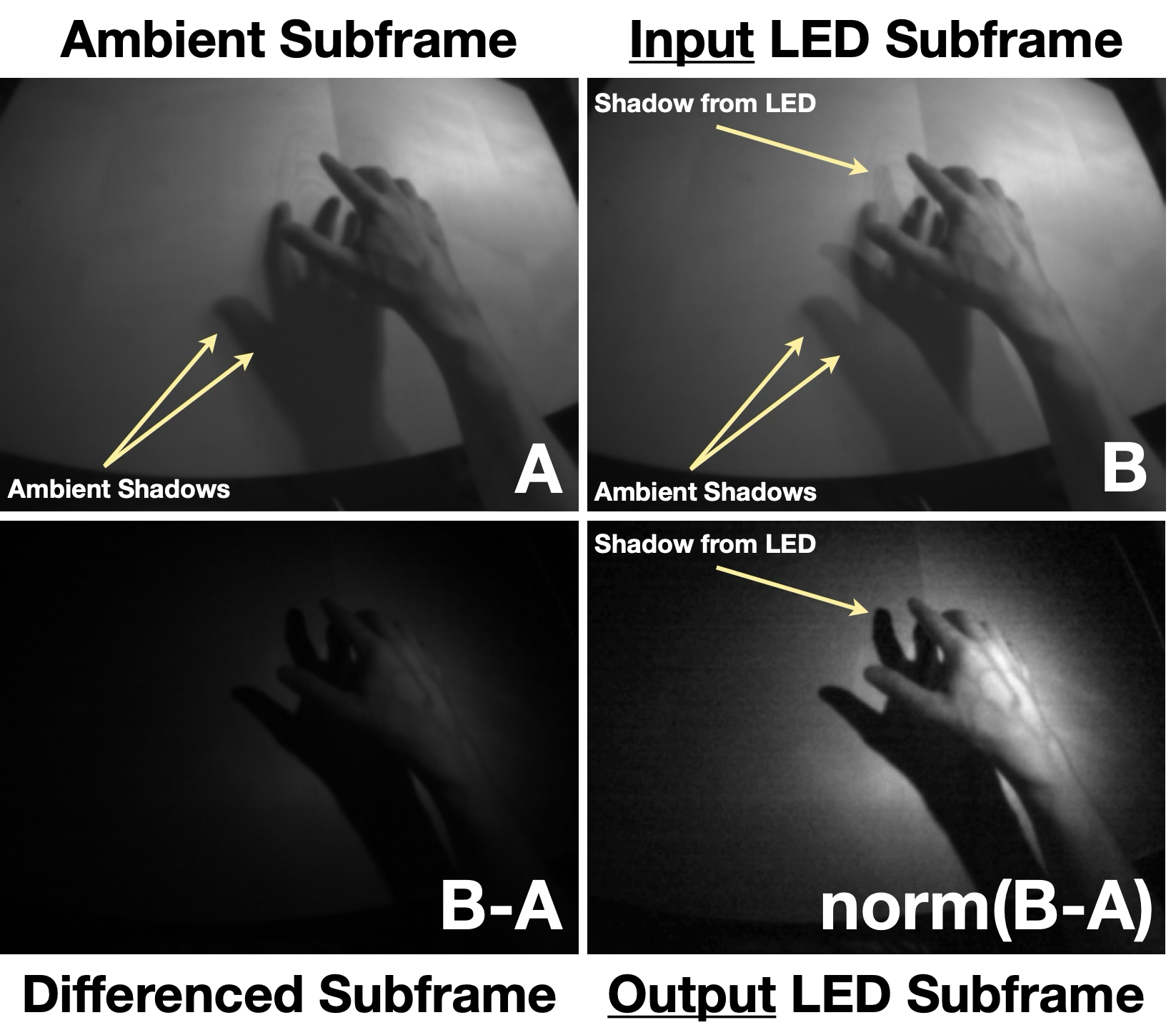}
     \caption{Overview of our shadow suppression process. Our ambient subframe (A) captures shadows cast by ambient light sources. When an illuminator is on (LED 3 in this example), we cast a new additional shadow into the scene (B). Note this shadow can be weaker than ambient shadows, as is the case in this example (see also Figure \ref{fig:shadowsuppression}). We then subtract the ambient frame from the LED-illuminated subframe (B-A). Having now subtracted ambient light, the frame becomes darker, and represents light only cast by the headset LED. To compensate for variable scene brightness (e.g., varying surface albedo and hand distance) we perform a final normalization --- norm(B-A) --- which accentuates the shadow. Note only a single shadow remains, the one cast by the headset LED. }
     \label{fig:shadowsuppressionprocess}
 \end{figure}

Importantly, light intensity is additive on image sensors (i.e., each CMOS pixel measures accumulated light, and if two or more light sources are contributing photons to this pixel, it will simply be the sum of intensities). We can use this property to great effect for removing unwanted shadows. Specifically, we can take our "no LEDs on" subframe, which captures any shadows generated from extraneous light sources, and simply subtract this from our other four subframes. This has the effect of removing the contribution of ambient light on those subframes, leaving only the illumination from that specific subframe's LED. An illustration of this process is shown in Figure \ref{fig:shadowsuppressionprocess}, with example outputs across various environmental conditions seen in Figure \ref{fig:shadowsuppression}.  

\subsection{Finger Tracking}

With extraneous shadows removed, we next move to track the hands in front of the user. In an integrated system, this information would already be available from the hand tracking provided by the headset software. However, as we are using our own camera, we cannot simply use the Quest 3's hand tracking result, and instead must compute our own. For this, we utilize our LED 1 subframe, which provides an illuminated view of the hand with minimal shadows (as LED 1 is almost directly inline with the camera). We run Google's MediaPipe hand tracker \cite{zhang_mediapipe_2020}, which provides 21 2.5D hand keypoints, though we only use the five fingertip points. We create 64x64 pixel patches centered on each fingertip. We use the wrist and Metacarpophalangeal (MCP) hand joints to normalize the size of the fingers (i.e., scale, irrespective of distance from the camera) scale, and then the  MCP and Proximal interphalangeal (PIP) finger joints to normalize the rotation (so that all fingers are pointing upwards in the patch). These finger patches are then passed to our ML model, described next.

\subsection{Machine Learning}
Our deep learning model takes in as input a finger patch and a finger ID and jointly predicts touch state and hover distance. Our model is a hybrid vision transformer, built on top of the FastViT T8 \cite{vasu_fastvit_2023} backbone. Finger patches contain contain N channels ($N \in {1, 2, 3, 4}$), one for each of the illuminator subframes. Finger patches are first passed through the backbone to produce image embeddings of size 768. These embeddings are then concatenated with a 5 dimensional one-hot encoded vector of the finger ID (thumb=1, pinky=5) to produce an embedding of size 773. Finally, this embedding is passed through a multi-layer perceptron (2 layers, hidden dimension 128, GeLU \cite{hendrycks_gaussian_2023} activations) that encodes the scaled distance of the finger to the surface. Touch state is obtained by sigmoid activating and thresholding this value. We smooth touch and hover distance predictions with a mean filter over the 30 most recent frames.

\subsection{Model Training Protocol}
In our subsequent user study, we employ a leave-one-participant-out cross validation scheme to train and evaluate our models. Our models were trained using the PyTorch and PyTorch Lightning deep learning frameworks. We initialized the FastViT backbone with ImageNet pretrained weights. We first trained the model to estimate touch state (by minimizing binary cross-entropy loss) and later fine-tuned the model to encode distance in the final logit (by minimizing the sum of mean absolute error and mean squared error). Models were trained for 10 epochs using the Adam optimizer, a batch size of 128 and a learning rate of 0.000003, which took about two hours on an NVIDIA 2080 Ti GPU. 
\begin{figure}[b]
    \centering
    \includegraphics[width=\linewidth]{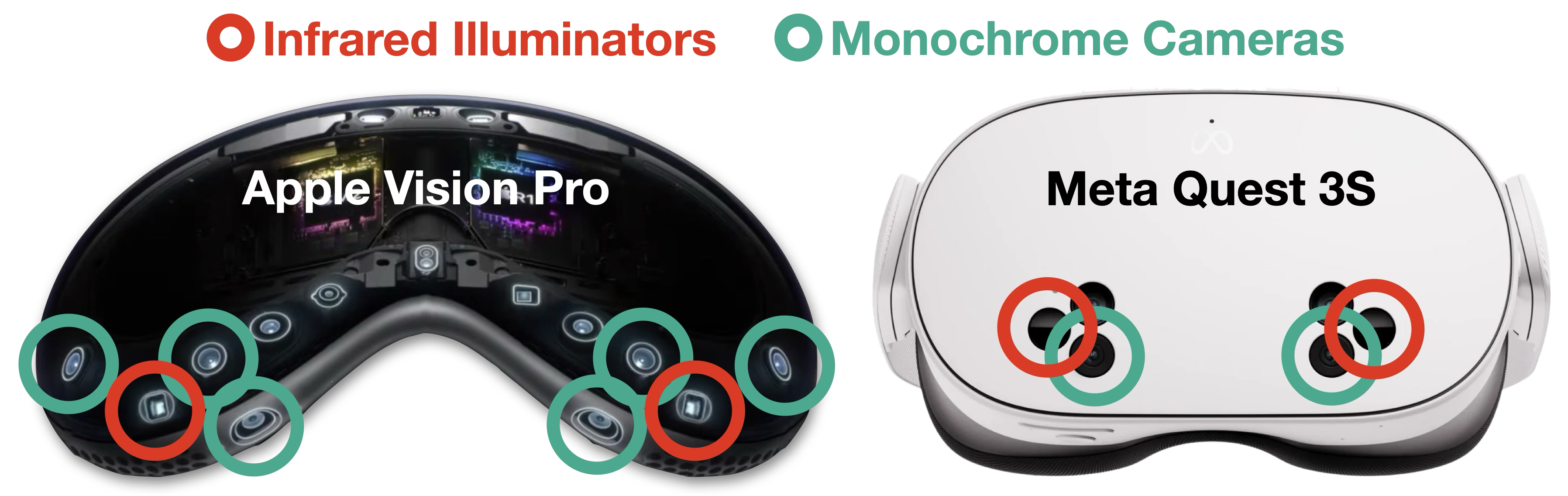}
    \caption{XR headsets on the market already include integrated infrared illuminators and infrared-sensitive cameras, suggesting EclipseTouch could be enabled via a software update. Additionally, we note that future AR glasses could also incorporate a single camera and LED in opposite corners of the frame, much like Ray-Ban Meta AI Glasses do today.}
    \label{fig:modernHeadsets}
\end{figure}

\subsection{Compute and Power Consumption}
\label{sec:computeAndPowerConsumption}
We carefully designed our model to be able to run as a lightweight background process, concurrent with numerous other models that already run on XR headsets (hand/body tracking, SLAM, etc). After training, we re-parameterize the model \cite{vasu_fastvit_2023} to an equivalent one with fewer parameters (total 3.3M parameters). On an M2 Macbook Air --- with similar hardware to the Apple Vision Pro --- our model has an inference time of 0.47~ms. This means that our model can potentially run at \textasciitilde2000 FPS, or run at e.g., 60 FPS consuming a small fraction of the headset's processing power.

As noted previously, our infrared LED illuminators consume 1.1~W when active. Only one LED is active at a time, and no LEDs are active 1/5th of the time, yielding a mean power draw of 0.9~W for illumination. Our microcontroller and LED driver board consumes 0.3~W. Our camera, running at 400~FPS, draws 0.8~W. As one reference point, the Meta Quest 3 draws \textasciitilde8.6 W of power during use (giving its 18.9 Wh battery a stated 2.2 hours of runtime).

\begin{figure*}[t]
    \centering
    \includegraphics[width=\linewidth]{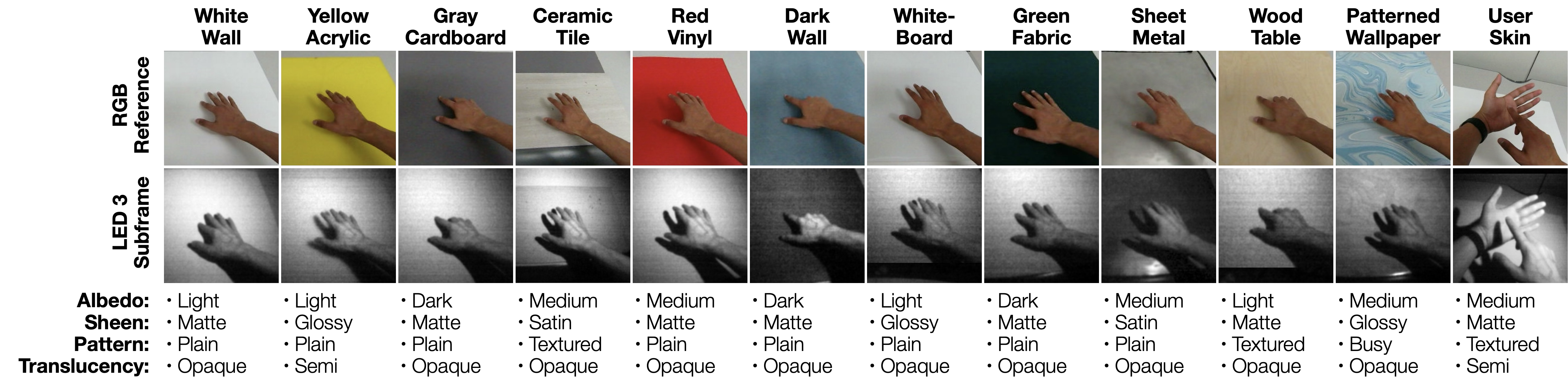}
    \caption{The twelve materials we tested in our study. Note the difference in appearance in RGB vs. infrared color spaces.}
    \label{fig:materialstested}
\end{figure*}

\subsection{Compatibility with Existing Headsets}\label{sec:compatibilitywithcontemporaryheadsets}
We note that several popular XR headsets already contain the requisite hardware to enable \systemname. For example, the Apple Vision Pro contains two infrared illuminators and six monochromatic cameras sensitive to infrared light (Figure \ref{fig:modernHeadsets}, left). The low-cost Meta Quest 3S contains two infrared illuminators and two monochromatic cameras (Figure \ref{fig:modernHeadsets}, right). In both cases, the illuminators are used to boost hand tracking performance in low-light conditions \cite{heaney_quest_2024}. However, their arrangement is also perfect for \systemname --- two infrared illuminators offset from one or more infrared cameras (more discussion in Section~\ref{ref:ledablation}). Thus, it is likely that \systemname could be enabled with a software update (at present, neither headset provides ad hoc surface touch segmentation, other than in a rough way using the native hand tracking).

\section{Data Collection Protocol}

To collect data to train and evaluate \systemname, we recruited 10 participants (4 female, 6 male, all right-handed) for a one-hour user study. Participants were compensated \$20 for their time. After completing consent paperwork, participants were fitted with our \systemname-instrumented Quest 3, which streamed our 80 FPS video composite to a local desktop via USB where it was saved. The study was conducted in a windowless room with controlled lighting. We measured lux, reported later, using a Sper Scientific 840022 Light Meter. During the study, participants stood in front of a standing desk or a wall. Unlike prior work, we did not restrict participants head motion or distance to surfaces. 

\begin{figure}[b]
    \centering
    \includegraphics[width=\linewidth]{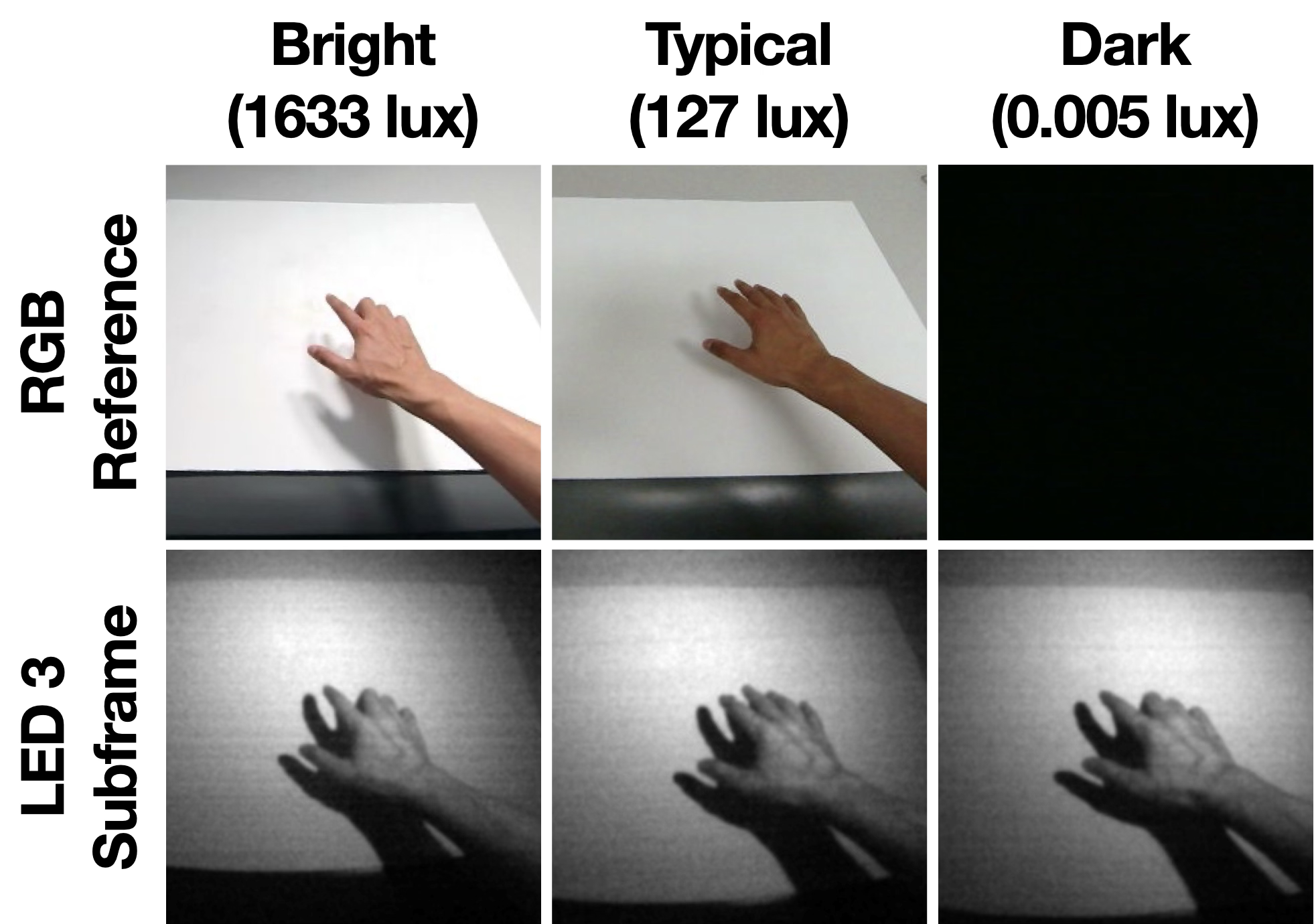}
    \caption{We tested three lighting conditions in our study: bright, typical, and dark. For reference only, we show the scene as it appears to a standard RGB camera along the top row. The bottom row shows the LED 3 illuminated subframe at the end of our pipeline (not shown are LED 1, 2 and 4 subframes, but the results are equivalent). In short, \systemname is reasonably agnostic to ambient lighting condition as it provides it own illumination. Note also how shadows cast from ambient sources are not visible in the processed frame due to shadow suppression.}
    \label{fig:lightingconditions}
\end{figure}

We designed our study to capture a variety of conditions such that we could later analyze system performance across different materials (n=12, of varying albedo, sheen, patterns and translucency; Figure \ref{fig:materialstested}), surface orientations (horizontal and vertical), and lighting conditions (bright, typical, and dark; Figure \ref{fig:lightingconditions}). It was not possible to fully cross these conditions due to combinatorial explosion, and so we designed blocks of sessions to collect data, varying a single experimental factor at time. 

All of the individual sessions followed the same basic data collection procedure. First, participants were instructed to continuously touch and drag on a presented surface with their index finger, during which time 30 seconds of data was recorded. This was immediately followed by a second session capturing 30 seconds of data in which participants were asked to hover their finger above the surface and to perform repeated in-air taps. These trials provided positive and negative examples to train our machine learning model. 

To collect data across a variety of surfaces, we curated a set of 11 diverse materials --- including wood, plastic, metal, fabric, and painted/printed surfaces --- seen in Figure~\ref{fig:materialstested} (along with a summary of their properties). These material samples were all cut down to 60$\times$40 cm so as to be easily swapped and organized during the study. Both touch and hover sessions of data were collected for all 11 materials, in a random presentation order, in a horizontal orientation (placed on a desk), and in typical home lighting (measured at 127 lux, with typical home illumination around 100-200 lux \cite{noauthor_lux_2025}). 

\begin{figure*}[t]
    \centering
    \includegraphics[width=\linewidth]{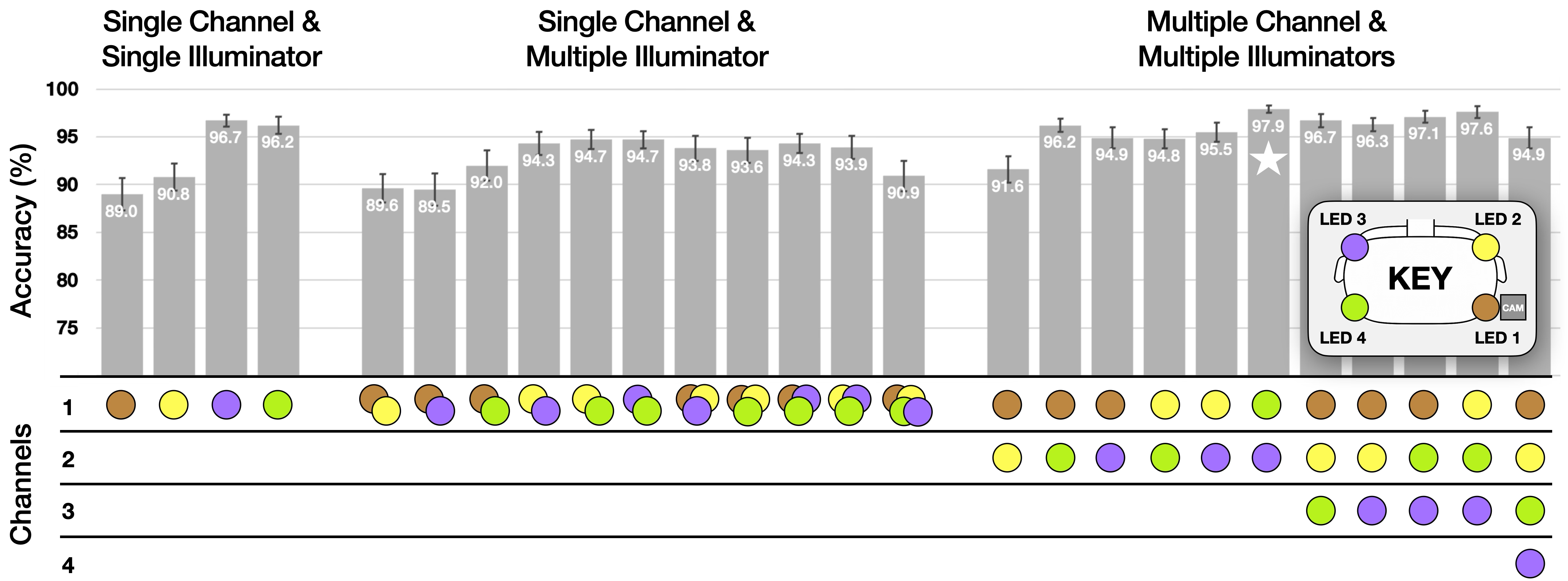}
    \caption{Results across illuminator combinations. A colored dot indicates that the corresponding illuminator is active in that channel. The best combination (LED 3 \& LED 4, operating in separate channels) is marked with a star.}
    \label{fig:acc_vs_led_combo}
\end{figure*}

Next, we collected data in two additional lighting conditions: bright (1633 lux) and dark (0.005 lux). For reference, typical TV studio lighting is around 1000 lux and a quarter moon on a cloudless night is 0.01 lux \cite{noauthor_lux_2025}. For this study block, we reduced our material set down to two: white wall and patterned wallpaper. The former was a representative plain and light-colored material, and the latter served as a more challenging surface, being darker, glossy, and patterned. For these two materials, we collected two additional sessions of data in each lighting condition and in a horizontal orientation. Note in the previous block of sessions, we already captured data for these two materials under typical lighting (127 lux). For our three lighting conditions, we provide reference visible-light photos and infrared-illuminated subframes in Figure \ref{fig:lightingconditions}. We also collected data for surfaces in a vertical orientation. For this, we use our same pared-down material set --- white wall and patterned wallpaper --- and our typical home lighting condition. 

Finally, the user's skin is a special, high-value input surface with unique optical properties (most notably subsurface scattering, creating diffuse shadows). In order to evaluate \systemname's ability to extend to skin input (our 12th material), we collected eight sessions of data on the palm: two sessions each with the hand held horizontal in our bright/typical/dark lighting conditions, and two sessions with the hand held vertically in typical lighting. 

In total, each participant completed 42 sessions of data collection. With each session producing 30 seconds of data and a system framerate of 80 Hz, multiplied by 10 participants, this process yielded 1,008,000 individual frames for training and analysis.

\section{Results and Discussion}\label{sec:results}
Our system prototype and study procedure was purposely designed to enable investigation of several important factors. First and foremost, we ran an ablation study to identify the most promising illuminator geometry, balancing accuracy with practicality. All subsequent results use this arrangement. Input to the skin is broken out as a special discussion. We conclude the section with two supplemental studies: hover distance estimation and multitouch. For touch prediction, we report classification accuracy and for hover distance estimation, we report mean absolute error in millimeters. Other than our illumination geometry ablation study (the next section), all results reported in this section are trained and tested with leave-one-participant-out cross validation, with no user or surface calibration. In all experiments, we always test on unseen users. 

Across all materials (including skin), in both orientations and all three lighting conditions, \systemname achieves an overall mean accuracy of 98.0\% (SD=0.3\%) using its best-performing LED 3 \& 4 illumination geometry. 

\subsection{Across Illumination Geometries}\label{ref:ledablation}
First, we evaluate the performance of \systemname under different illuminator configurations. Our prototype hardware has four illuminators (Figure~\ref{fig:pipelineandhardware}), which leads to 15 possible combinations ($2^{4} - 1$) for illuminator placement on the headset. To enable triggerless operation and higher framerates, illuminators could also be turned on in groups of two, three, or four, and so we additionally ablate these 11 extra combinations. Due to the additive property of light transport, we can simulate these combinations by averaging the subframes from each of the LEDs. 
Figure~\ref{fig:acc_vs_led_combo} shows all the 26 illuminator combinations we tested. Prior to training each model, we modify the model architecture to accept images with different numbers of channels, by modifying the input channels of the first convolution layer of the backbone.
We used six participants' data for training, and the remaining four participants' data for testing. 

Results from this ablation study are shown in Figure~\ref{fig:acc_vs_led_combo}. 
Focusing first on the Single Channel, Single Illuminator results, it is clear that LED illuminators 3 and 4 yield the best results (96.7\% and 96.2\% accuracy).
Indeed, there is a clear pattern in performance relating to illuminator-camera distance.
Figure~\ref{fig:pipelineandhardware} shows shadows formed by each of these illuminators. We note that illuminators 3 and 4 are the furthest offset from the camera and produce the most prominent shadows of the touching finger. Next farthest from the camera is illuminator 2, which produces notably smaller shadows of the touching finger. Finally illuminator 1, which is in-line with the camera, produces almost no shadow, leading to the worst performance of all the illuminators (89.0\%).

Moving on to Single Channel, Multiple Illuminator configurations, where multiple illuminators are on at the same time, we observe that the best combinations are LED 3 + 4 (94.7\%), as well as LED 2 + 4 (also 94.7\%). While this performance is high, we observe that all combinations in this category consistently perform worse than their individual channel counterparts, as well as the best performing single-illuminator configurations. Simultaneously turning on multiple illuminators had no performance benefit over using one or more LEDs at optimal locations.

Finally, looking at Multiple Channel, Multiple Illuminator results, we find the best-performing combination is LED 3 \& 4 at 97.9\%, followed closely by LED 2, 3 \& 4 at 97.6\%. This makes sense, especially considering the fact that LED illuminators 3 and 4 already achieved high accuracies on their own. 

\begin{figure}[t]
    \centering
    \includegraphics[width=\linewidth]{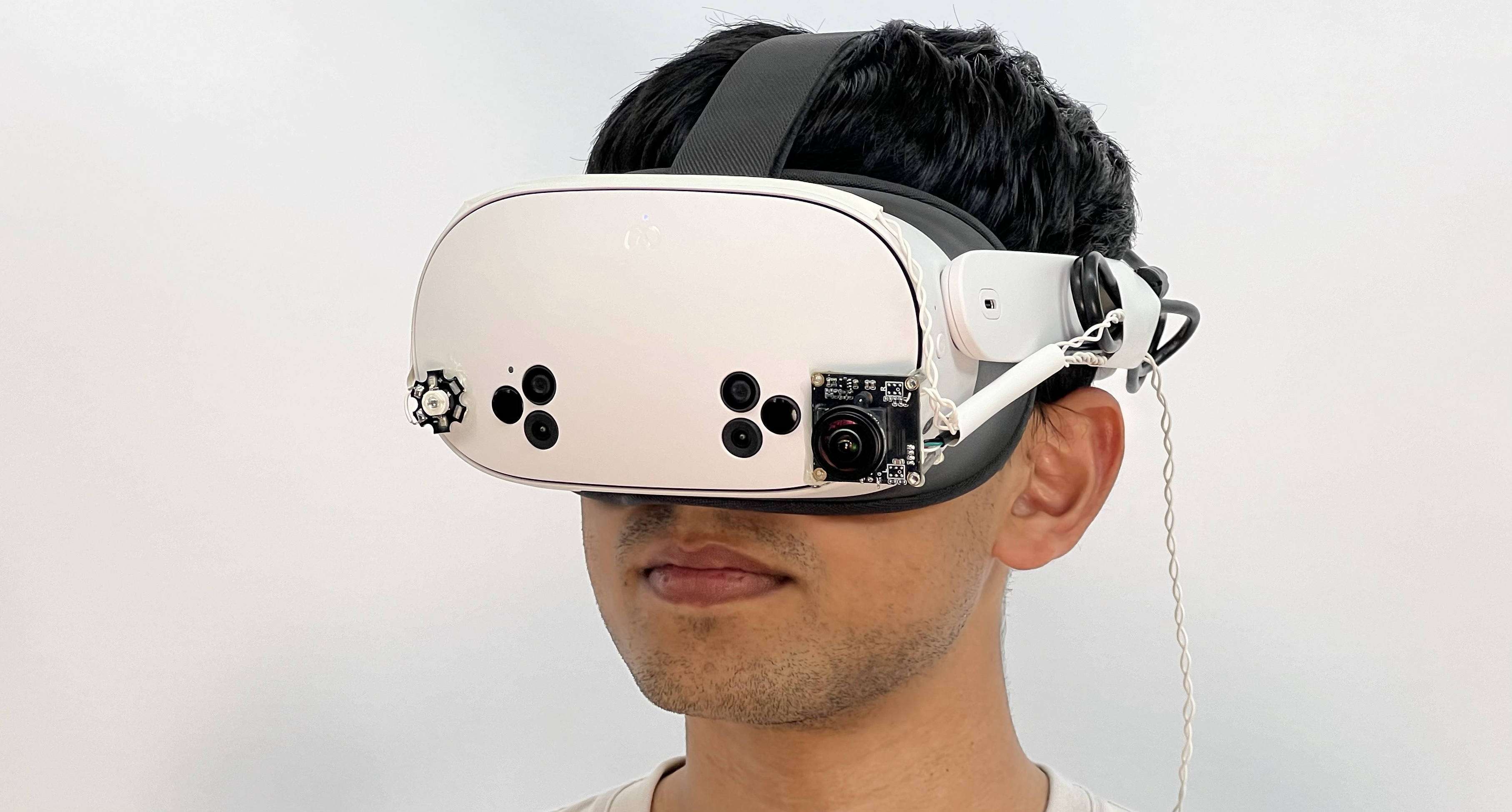}
    \caption{Based on our evaluation results, we created a final prototype featuring a single infrared emitter and camera. }
    \label{fig:new_prototype}
\end{figure}

Notably, the accuracy with even just one LED at an optimal location is similar to the best performing combination (96.7\% for LED 3 alone vs. 97.9\% for LEDs 3 \& 4). This is promising, as existing headsets already include a pair of illuminators offset from integrated cameras (see Figure \ref{fig:modernHeadsets} and Section~\ref{sec:compatibilitywithcontemporaryheadsets}), potentially allowing \systemname to be enabled through a software update. We also created a final prototype for demo purposes, seen in Figure \ref{fig:new_prototype}, which features a single illuminator based on these results. For all subsequent study results, we report performance on this best-performing, multi-channel illuminator configuration: LED 3 \& 4.

\begin{figure}[b]
    \centering
    \includegraphics[width=\linewidth]{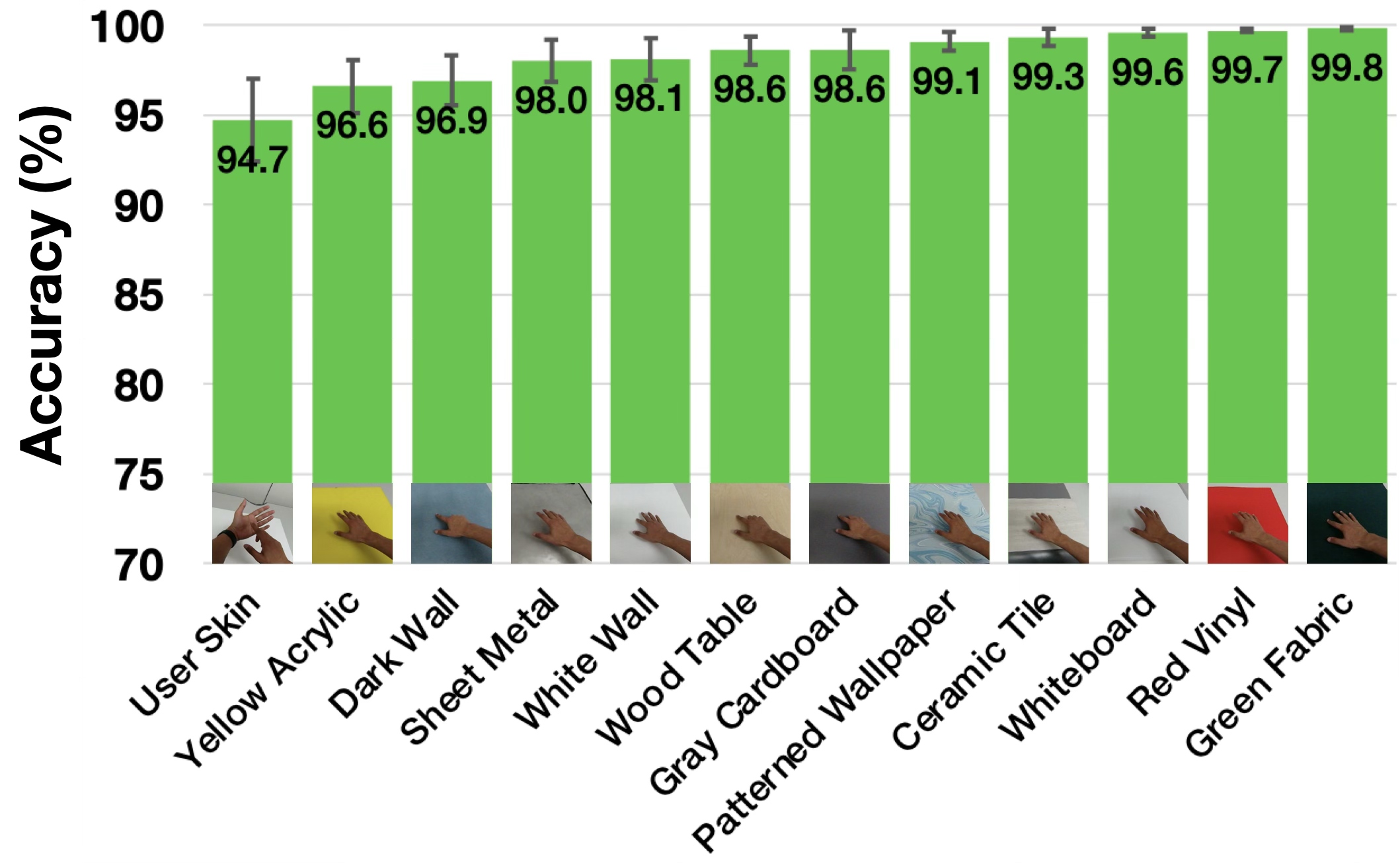}
    \caption{Touch classification accuracy vs. surface material.}
    \label{fig:accvsmaterial}
\end{figure}

\subsection{Across Surface Materials}
A key component of our study was to evaluate the performance of \systemname on a variety of everyday materials (see Figure~\ref{fig:materialstested} for the 11 non-skin materials that we tested). Importantly, these materials had a wide range of properties, including varying albedo, sheen, patterns and translucency. Results from this evaluation are shown in Figure~\ref{fig:accvsmaterial}. Across all materials tested, performance remains high, averaging 98.6\% accuracy. This ranges from 96.6\% for the yellow acrylic material to 99.8\% for the green fabric material. Note that 9 out of 11 materials tested have accuracies above 98.0\%, indicating strong generalization of \systemname across surface materials. 

Performance remains similar across albedo types, with light materials averaging 98.2\% (SD=1.3\%), medium materials averaging 99.0\% (SD=0.7\%), and dark materials averaging 98.4\% (SD=1.4\%) accuracy. This is encouraging, since darker materials often pose a challenge to other vision-based systems.

Similarly, performance remains comparable across sheen types, with matte materials averaging 98.6\% (SD=1.1\%), satin materials averaging 98.7\% (SD=0.9\%), and glossy materials averaging 98.4\% (SD=1.6\%). Note that while we do include glossy materials, none of our materials were fully reflective, like glass. These materials do not produce shadows and thus cannot be supported. 

Also encouraging is that performance remains high with materials with busy patterns; over 99.0\% accuracy for our patterned wallpaper, wood table, and ceramic tile surfaces. 

The worst performing material was the yellow acrylic, which had a mean accuracy of 96.6\%. This material is semi-translucent, and produces shadows with softer edges as compared to the other materials in our study, potentially reducing performance. Opaque materials averaged 98.8\% accuracy (SD=0.9\%). 

\subsection{Across Lighting Conditions}
\begin{figure}[b]
    \centering
    \includegraphics[width=1.0\linewidth]{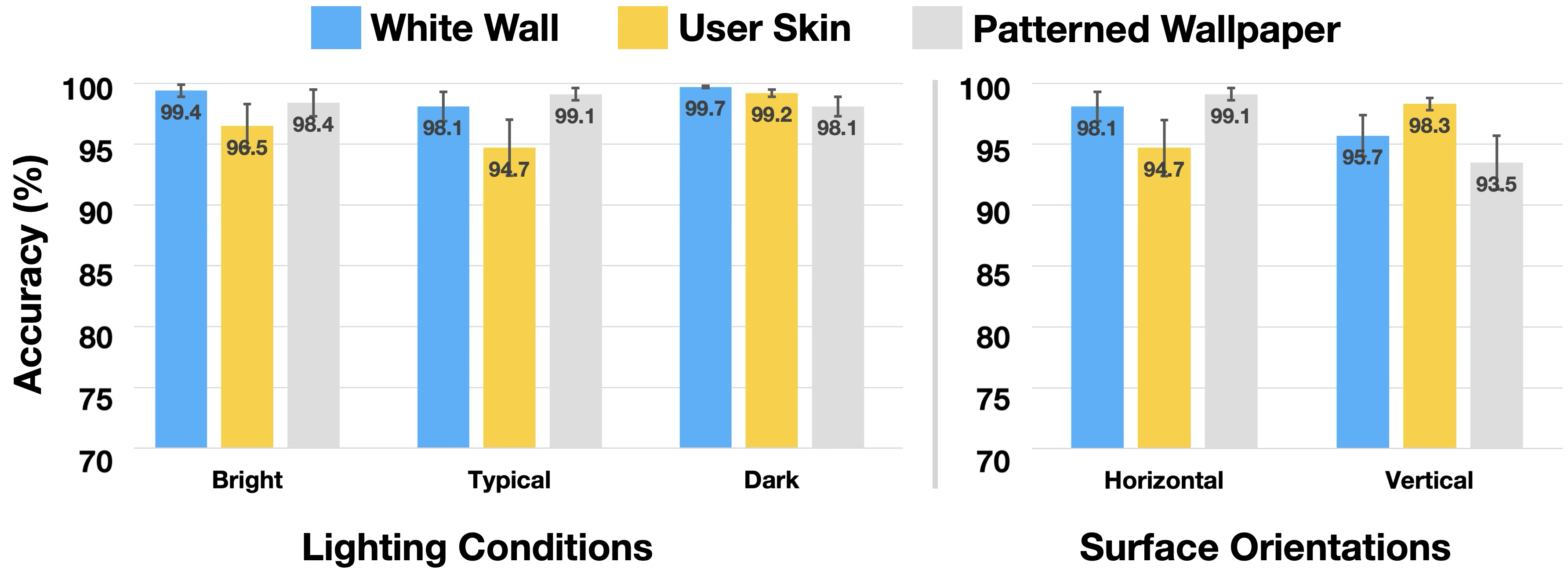}
    \caption{Touch classification accuracy vs. lighting condition (left) and surface orientation (right).}
    \label{fig:accvslightingandorientation}
\end{figure}

Next, we analyze performance across lighting conditions. Figure~\ref{fig:lightingconditions} shows the three lighting conditions we tested in this study, ranging from dark (0.005 lux) to bright (1633 lux) lighting. Results for the three materials we studied across lighting conditions can be seen in Figure~\ref{fig:accvslightingandorientation}. On average, touch accuracies were 98.1\% (SD=1.5\%) for the bright condition, 97.3\% (SD=2.3\%) for the typical condition, and 99.0\% (SD=0.8\%) for the dark condition. Overall, performance remains similar across lighting conditions, with a slight increase in performance in the darkest lighting condition. This is an encouraging result, since most prior work either did not test or did not work well across such a wide range of lighting conditions. 

\subsection{Across Surface Orientations}
Touch performance across the two orientations we tested can be seen in Figure~\ref{fig:accvslightingandorientation}. On average, performance was slightly better for the horizontal condition (97.3\% accuracy, SD=2.3\%) vs. the vertical condition (95.8\% accuracy, SD=2.4\%). Notably, the patterned wallpaper material had a sharp decrease in performance from 99.1\% to 93.5\% accuracy. We note that in our user study, participants typically had their hand closer to the headset in the vertical condition vs. the horizontal condition, which could have contributed to this decrease in performance. Furthermore, shadows of the touching finger look different in the vertical condition as compared to the horizontal condition, and since the majority of our training data was collected in the horizontal orientation, we hypothesize our models may be slightly biased towards this orientation. 

\begin{figure*}[t]
    \centering
    \begin{minipage}[t]{.48\textwidth}
        \centering
        \includegraphics[width=\linewidth]{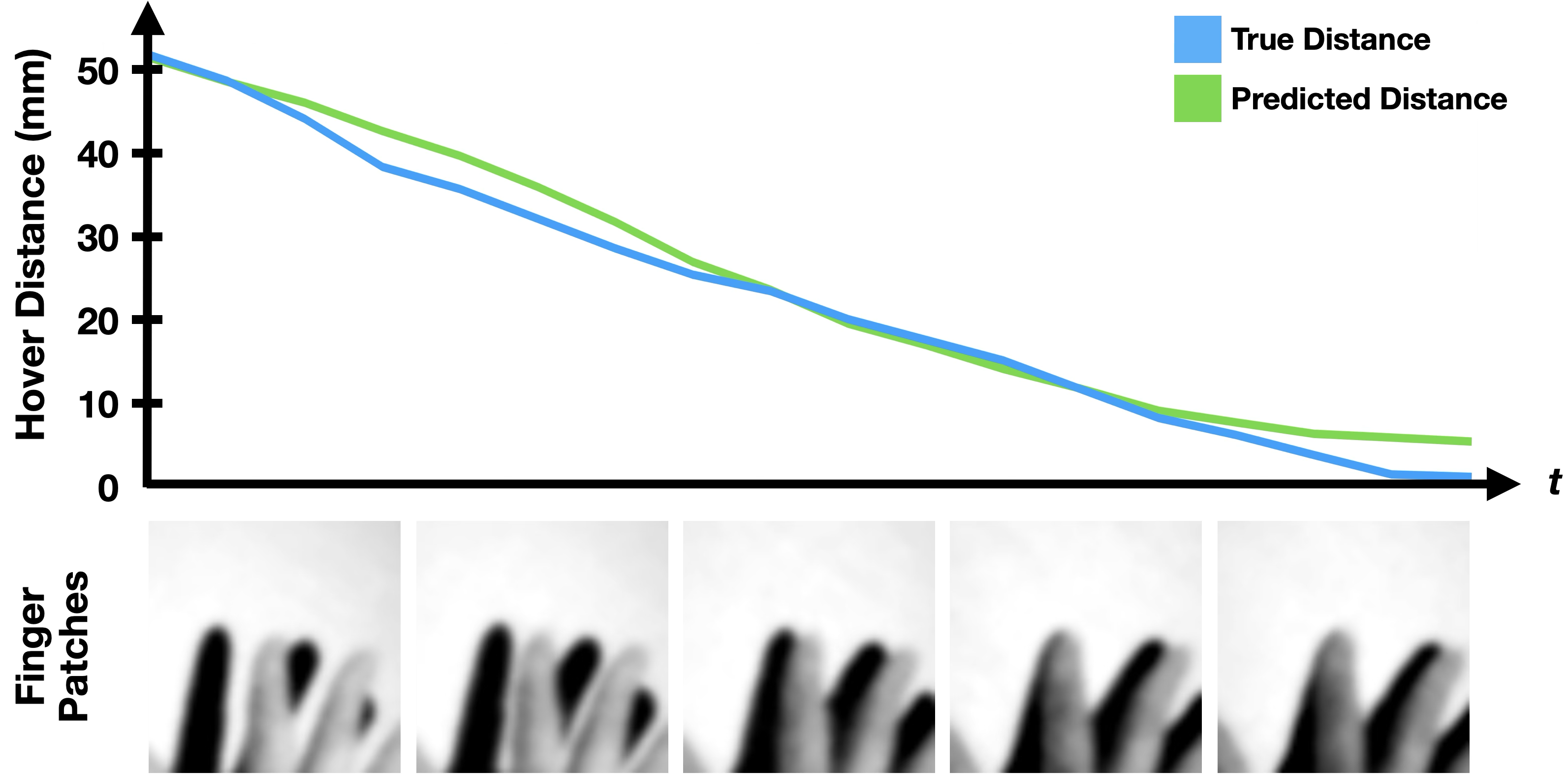}
    \caption{Example 250 ms sequence of an index finger descending to tap a surface. Note how the shadow evolves over time (illuminated by LED 3), converging towards the finger, and essentially disappearing upon contact. Our model uses this visual information to predict hover distance.}
    \label{fig:distanceestimationexample}
    \end{minipage}\hfill
    \begin{minipage}[t]{.48\textwidth}
        \centering
            \includegraphics[width=\linewidth]{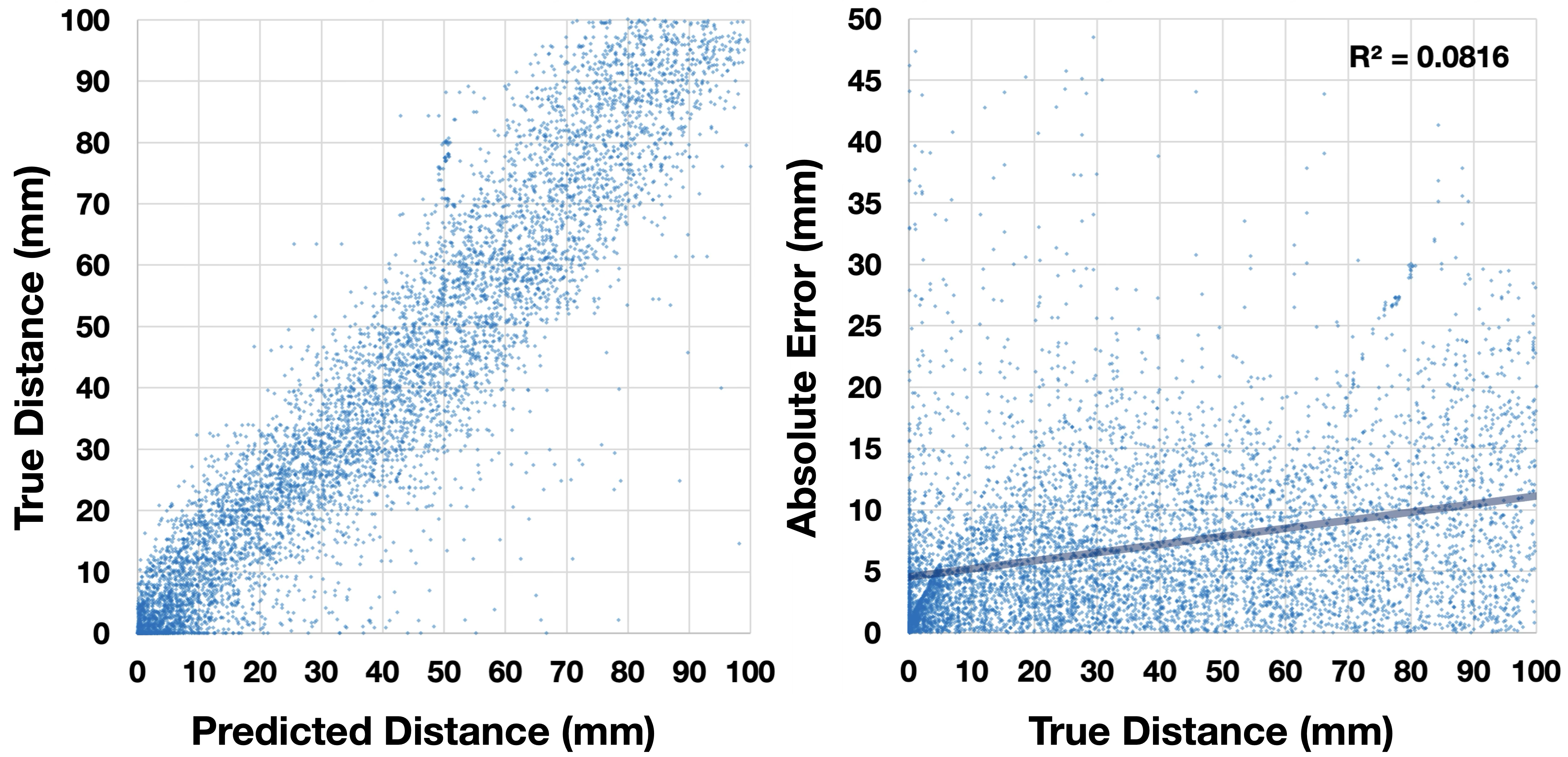}
    \caption{Left: Predicted vs. true hover distance. Overall, \systemname is fairly linear and accurate, up to about 10~cm (mean error of 6.9~mm). Right: Mean absolute hover distance error vs. true distance --- note the slight upward trend in error. }
    \label{fig:distanceresults}
    \end{minipage}   
\end{figure*}

\subsection{On-Skin Touch Detection}\label{sec:onskintouch}
An important surface material to study for touch input is the user's skin, as it is always available for input on the go. However, skin has unique properties that set it apart from the other materials we tested. It has varying albedo, surface texture, and some translucency (leading to subsurface scattering of light). It is also deformable and non-planar. For this reason, we separately trained a model for skin input. We envision this model being used when the XR headset's existing hand tracking detects probable hand-to-skin input.

On-skin touch detection results are shown in Figures~\ref{fig:accvsmaterial} and \ref{fig:accvslightingandorientation}. Compared to other materials, detection accuracy was slightly lower, at 94.7\% accuracy (SD=2.3\%). We note, however, that this performance is still competitive with prior work \cite{mollyn_egotouch_2024, harrison_omnitouch_2011}. Across lighting conditions, performance also remains high, with a noticeable jump in performance in the dark (99.2\% accuracy). We note that prior vision-based on-skin touch systems, such as our own EgoTouch \cite{mollyn_egotouch_2024}, has essentially 0\% accuracy in the dark. Across orientations, performance is higher in the vertical orientation (98.3\%) vs. the horizontal orientation (94.7\%). 

\subsection{Supplemental Study: Hover Distance}
As a user moves their finger towards a surface, the shadow cast by that finger moves and the distance between the fingertip and the tip of the shadow changes proportionally (Figure~\ref{fig:distanceestimationexample}). For this reason, we hypothesized that it would be possible to train a model to estimate finger hover distance above a surface. To train and evaluate this model, we ran a small supplemental user study with 5 participants (2 male, 3 female). Participants filled out consent paperwork and were fitted with the \systemname prototype. Participants stood in front of a table in a typically lit room. To collect ground truth distance of the finger from the surface, we affixed a HD USB webcam to the side of the table, so as to track the participant's touching finger from the side. At the start of the study, the experimenter calibrated the pixel displacements of the user's hand in the webcam view to real world units (mm). Then, participants were asked to lift their index finger up and down above a certain location on the surface. Participants were instructed to only lift their finger vertically, so as to accurately track their fingertip's ground truth distance from the surface. 
We collected three sessions lasting 60 seconds each, for a total of 72,000 frames at 80 FPS.

To train this model, we simply fine-tuned our existing \systemname model to directly estimate hover distance in the output touch classification logit (before activation). The main motivation for this design was that hover distance and touch classification confidence are highly correlated. Furthermore, this allowed us to have a single unified model that estimated both hover distance and touch state simultaneously. All models were trained with a leave-one-participant-out cross validation scheme. 

Results from this supplemental study can be seen in Figure~\ref{fig:distanceresults}. In our hover distance region of interest, below 10 cm, our model had a mean absolute error of 6.9~mm. For ground truth distances less than 1~cm away from the surface, our model had a mean absolute error of 2.5~mm. Furthermore, we note that error increases slightly as distance from the surface increases (Figure~\ref{fig:distanceresults}). Example predictions of this model as a user touches a surface can be seen in Figure~\ref{fig:distanceestimationexample}.

\begin{figure}[b]
    \centering
    \includegraphics[width=\linewidth]{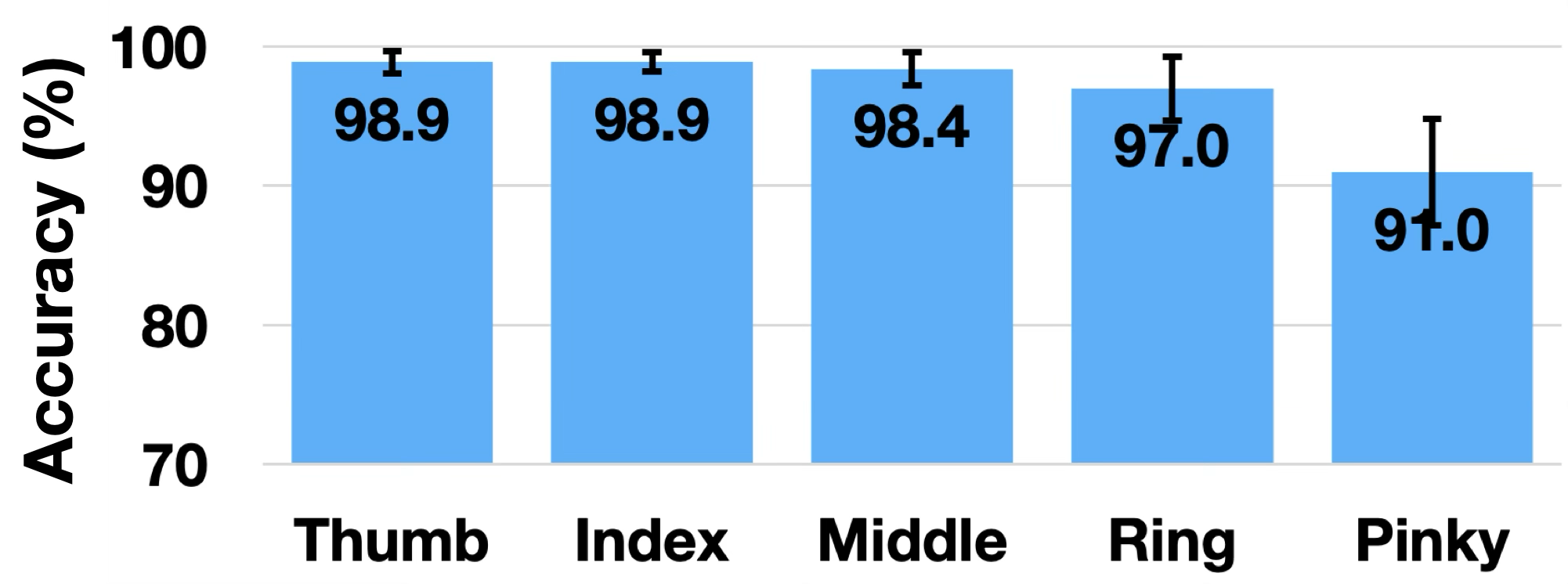}
    \caption{Results from our multitouch supplementary study, broken out by inputting finger. }
    \label{fig:accvsfinger}
\end{figure}
\subsection{Supplemental Study: Multitouch}
To evaluate the performance of \systemname on multiple fingers, we ran a small supplemental study (in tandem with the hover distance estimation study). For this study, we collected four types of sessions, all with the white wall material in typical lighting. First, participants were asked to touch and drag across the surface with all their fingers simultaneously. Next, participants were asked to hover close to the surfaces and to perform in-air taps with all their fingers. Third, participants were asked to perform a pinch and zoom motion with their index and thumb fingers while touching the surface. Finally, participants were asked to perform the pinch and zoom motion in the air while hovering close to the surface. Each session lasted 30 seconds, and we collected three rounds per session, for a total of 144,000 frames at 80 FPS.
We then fine-tuned our models with this data. Similar to before, we employed a leave-one-participant-out cross validation scheme to train our models.

Results from this study are shown in Figure \ref{fig:accvsfinger}. Similar to \cite{liang_shadowtouch_2023}, we observe that the model performs similarly across the thumb, index and middle fingers (98.7\% accuracy, SD=0.3\%). Performance drops slightly for the ring finger (97\%) and drops further for the pinky finger (91\%), as they are sometimes not visible to the camera.

\section{Limitations \& Future Work}
\label{sec:limitations}
While \systemname represents a useful and practical increment over prior work, there is still room for improvement to achieve touchscreen-like performance. Foremost, similar to other vision-based approaches, \systemname needs line-of-sight to the touching finger to function. On modern headsets, this issue has been partially addressed by incorporating multiple cameras on the headset, expanding the field of view outside that of the users. 

We also note that \systemname does not work across all surface materials. In particular, highly-reflective and transparent materials will fail, such as mirrors and glass (as no shadows are visible). In future work, the reflection of the finger on the surface itself could be used to estimate touch contact \cite{pak-kiu_chung_mirrortrack_2008}. Apart from this, we also observed that very dark materials (in infrared) did not cast shadows, nor did highly 3D-textured surfaces (e.g., piled carpets, fur, fluffy clothing). Across all materials tested in our study, the two worst-performing were User Skin and Yellow Acrylic (the only two semi-translucent materials we tested). One hypothesis for this reduced performance is that subsurface scattering interferes with cast shadows. In the future, \systemname could use its cameras and computer vision to identify unsuitable surfaces, and steer users to utilize compatible ones. 

The power draw of \systemname could be an obstacle to adoption in commercial systems (read more about the power consumption of our prototype in Section \ref{sec:computeAndPowerConsumption}). In general, active illumination is power expensive. In systems using light emitters, such as the LIDAR and TrueDepth sensors in Apple's iPhones, the sensor is only turned on when needed. For example, the auto-unlock feature is first triggered by a motion event detected by a lower-power IMU, before turning on the more power-expensive TrueDepth sensor for Face ID. Even when on, the duty cycle is kept very low. A similar approach could be used for \systemname. Contemporary XR headsets already build a model of the environment and track the user's hands for input. This data could be used to activate \systemname opportunistically, when the hands are closer than e.g., 30 cm to a surface. When active, the infrared LEDs could be strobed for very short durations (our cameras are already externally triggered, so synchronization is not an issue), further reducing power consumption. 

Another condition where \systemname does not presently function is in direct sunlight. As noted in our study, our bright lighting condition was 1633 lux (exceeding that of typical TV studio lighting at 1000 lux \cite{noauthor_lux_2025}). It was certainly bright compared to a typical office, but not bright compared to the direct Sun, which can be 100,000 lux \cite{noauthor_lux_2025}. At this level of illumination, our 1.1 W LEDs cannot compete, and the shadows they cast simply disappear into noise. Other active illumination sensing methods have employed various strategies to combat this, including modulating and polarizing light, as well as using very narrow bandpass optical filters. This is how, e.g., depth sensors such as Microsoft's Kinect and Apple's iPhone LIDAR can work in outdoor scenes. 

There are also opportunities to further generalize \systemname to new surfaces. For instance, we are exploring how game engines could be used to generate synthetic data, since the phenomenon of shadow casting is well developed for games. Alternatively, we could also use new foundation models for relighting \cite{iclight2025} and hand generation \cite{chen_foundhand_2024} to augment and create new synthetic data on a variety of materials. Future work could also explore touch detection on irregular surfaces, beyond the palm, as well as fusing multiple vision modalities (e.g. RGB \& IR).

\section{Conclusion}
We have presented \systemname, a new headset-only system for detecting touches on everyday surfaces, including the user's skin, using worn infrared shadow casting. Moving beyond prior work, our results show that this approach is quite accurate (98.0\% touch accuracy, 6.9 mm hover distance error), and works across a wide range of surface materials, lighting conditions and orientations, while running efficiently and with low latency. Our ablation studies reveal optimal illuminator arrangement geometries and suggest that \systemname could be implemented in existing headsets through a software update.


\bibliographystyle{ACM-Reference-Format}
\bibliography{references}


\begin{thebibliography}{76}


\ifx \showCODEN    \undefined \def \showCODEN     #1{\unskip}     \fi
\ifx \showDOI      \undefined \def \showDOI       #1{#1}\fi
\ifx \showISBNx    \undefined \def \showISBNx     #1{\unskip}     \fi
\ifx \showISBNxiii \undefined \def \showISBNxiii  #1{\unskip}     \fi
\ifx \showISSN     \undefined \def \showISSN      #1{\unskip}     \fi
\ifx \showLCCN     \undefined \def \showLCCN      #1{\unskip}     \fi
\ifx \shownote     \undefined \def \shownote      #1{#1}          \fi
\ifx \showarticletitle \undefined \def \showarticletitle #1{#1}   \fi
\ifx \showURL      \undefined \def \showURL       {\relax}        \fi
\providecommand\bibfield[2]{#2}
\providecommand\bibinfo[2]{#2}
\providecommand\natexlab[1]{#1}
\providecommand\showeprint[2][]{arXiv:#2}

\bibitem[Adajania et~al\mbox{.}(2010)]%
        {adajania_virtual_2010}
\bibfield{author}{\bibinfo{person}{Y. Adajania}, \bibinfo{person}{J. Gosalia}, \bibinfo{person}{A. Kanade}, \bibinfo{person}{H. Mehta}, {and} \bibinfo{person}{N. Shekokar}.} \bibinfo{year}{2010}\natexlab{}.
\newblock \showarticletitle{Virtual {Keyboard} {Using} {Shadow} {Analysis}}. In \bibinfo{booktitle}{\emph{2010 3rd {International} {Conference} on {Emerging} {Trends} in {Engineering} and {Technology}}}. \bibinfo{pages}{163--165}.
\newblock
\urldef\tempurl%
\url{https://doi.org/10.1109/ICETET.2010.115}
\showDOI{\tempurl}
\newblock
\shownote{ISSN: 2157-0485}.


\bibitem[Agarwal et~al\mbox{.}(2007)]%
        {agarwal_high_2007}
\bibfield{author}{\bibinfo{person}{Ankur Agarwal}, \bibinfo{person}{Shahram Izadi}, \bibinfo{person}{Manmohan Chandraker}, {and} \bibinfo{person}{Andrew Blake}.} \bibinfo{year}{2007}\natexlab{}.
\newblock \showarticletitle{High precision multi-touch sensing on surfaces using overhead cameras}. In \bibinfo{booktitle}{\emph{Second Annual IEEE International Workshop on Horizontal Interactive Human-Computer Systems (TABLETOP'07)}}. IEEE, \bibinfo{pages}{197--200}.
\newblock


\bibitem[Benko and Wilson(2009)]%
        {benko_depthtouch_2009}
\bibfield{author}{\bibinfo{person}{Hrvoje Benko} {and} \bibinfo{person}{Andrew Wilson}.} \bibinfo{year}{2009}\natexlab{}.
\newblock \showarticletitle{DepthTouch: Using depth-sensing camera to enable freehand interactions on and above the interactive surface}. In \bibinfo{booktitle}{\emph{Proceedings of the IEEE workshop on tabletops and interactive surfaces}}, Vol.~\bibinfo{volume}{8}. \bibinfo{pages}{21}.
\newblock


\bibitem[Chen et~al\mbox{.}(2025)]%
        {chen_foundhand_2024}
\bibfield{author}{\bibinfo{person}{Kefan Chen}, \bibinfo{person}{Chaerin Min}, \bibinfo{person}{Linguang Zhang}, \bibinfo{person}{Shreyas Hampali}, \bibinfo{person}{Cem Keskin}, {and} \bibinfo{person}{Srinath Sridhar}.} \bibinfo{year}{2025}\natexlab{}.
\newblock \showarticletitle{FoundHand: Large-Scale Domain-Specific Learning for Controllable Hand Image Generation}. In \bibinfo{booktitle}{\emph{Proceedings of the Computer Vision and Pattern Recognition Conference (CVPR)}}. \bibinfo{pages}{17448--17460}.
\newblock


\bibitem[Chen et~al\mbox{.}(2020)]%
        {chen_estimating_2020}
\bibfield{author}{\bibinfo{person}{Nutan Chen}, \bibinfo{person}{Göran Westling}, \bibinfo{person}{Benoni~B. Edin}, {and} \bibinfo{person}{Patrick Van Der~Smagt}.} \bibinfo{year}{2020}\natexlab{}.
\newblock \showarticletitle{Estimating {Fingertip} {Forces}, {Torques}, and {Local} {Curvatures} from {Fingernail} {Images}}.
\newblock \bibinfo{journal}{\emph{Robotica}} \bibinfo{volume}{38}, \bibinfo{number}{7} (\bibinfo{date}{July} \bibinfo{year}{2020}), \bibinfo{pages}{1242--1262}.
\newblock
\showISSN{0263-5747, 1469-8668}
\urldef\tempurl%
\url{https://doi.org/10.1017/S0263574719001383}
\showDOI{\tempurl}


\bibitem[Dupré et~al\mbox{.}(2024)]%
        {dupre_tripad_2024}
\bibfield{author}{\bibinfo{person}{Camille Dupré}, \bibinfo{person}{Caroline Appert}, \bibinfo{person}{Stéphanie Rey}, \bibinfo{person}{Houssem Saidi}, {and} \bibinfo{person}{Emmanuel Pietriga}.} \bibinfo{year}{2024}\natexlab{}.
\newblock \showarticletitle{{TriPad}: {Touch} {Input} in {AR} on {Ordinary} {Surfaces} with {Hand} {Tracking} {Only}}. In \bibinfo{booktitle}{\emph{Proceedings of the {CHI} {Conference} on {Human} {Factors} in {Computing} {Systems}}}. \bibinfo{publisher}{ACM}, \bibinfo{address}{Honolulu HI USA}, \bibinfo{pages}{1--18}.
\newblock
\showISBNx{9798400703300}
\urldef\tempurl%
\url{https://doi.org/10.1145/3613904.3642323}
\showDOI{\tempurl}


\bibitem[Fan and Xiao(2022)]%
        {fan_reducing_2022}
\bibfield{author}{\bibinfo{person}{Neil~Xu Fan} {and} \bibinfo{person}{Robert Xiao}.} \bibinfo{year}{2022}\natexlab{}.
\newblock \showarticletitle{Reducing the {Latency} of {Touch} {Tracking} on {Ad}-hoc {Surfaces}}.
\newblock \bibinfo{journal}{\emph{Proc. ACM Hum.-Comput. Interact.}} \bibinfo{volume}{6}, \bibinfo{number}{ISS} (\bibinfo{date}{Nov.} \bibinfo{year}{2022}), \bibinfo{pages}{577:489--577:499}.
\newblock
\urldef\tempurl%
\url{https://doi.org/10.1145/3567730}
\showDOI{\tempurl}


\bibitem[Funk et~al\mbox{.}(2015)]%
        {funk_interactive_2015}
\bibfield{author}{\bibinfo{person}{Markus Funk}, \bibinfo{person}{Stefan Schneegass}, \bibinfo{person}{Michael Behringer}, \bibinfo{person}{Niels Henze}, {and} \bibinfo{person}{Albrecht Schmidt}.} \bibinfo{year}{2015}\natexlab{}.
\newblock \showarticletitle{An {Interactive} {Curtain} for {Media} {Usage} in the {Shower}}. In \bibinfo{booktitle}{\emph{Proceedings of the 4th {International} {Symposium} on {Pervasive} {Displays}}}. \bibinfo{publisher}{ACM}, \bibinfo{address}{Saarbruecken Germany}, \bibinfo{pages}{225--231}.
\newblock
\showISBNx{978-1-4503-3608-6}
\urldef\tempurl%
\url{https://doi.org/10.1145/2757710.2757713}
\showDOI{\tempurl}


\bibitem[Gong et~al\mbox{.}(2020)]%
        {gong_acustico_2020}
\bibfield{author}{\bibinfo{person}{Jun Gong}, \bibinfo{person}{Aakar Gupta}, {and} \bibinfo{person}{Hrvoje Benko}.} \bibinfo{year}{2020}\natexlab{}.
\newblock \showarticletitle{Acustico: {Surface} {Tap} {Detection} and {Localization} using {Wrist}-based {Acoustic} {TDOA} {Sensing}}. In \bibinfo{booktitle}{\emph{Proceedings of the 33rd {Annual} {ACM} {Symposium} on {User} {Interface} {Software} and {Technology}}} \emph{(\bibinfo{series}{{UIST} '20})}. \bibinfo{publisher}{Association for Computing Machinery}, \bibinfo{address}{New York, NY, USA}, \bibinfo{pages}{406--419}.
\newblock
\showISBNx{978-1-4503-7514-6}
\urldef\tempurl%
\url{https://doi.org/10.1145/3379337.3415901}
\showDOI{\tempurl}


\bibitem[Grady et~al\mbox{.}(2024)]%
        {grady_pressurevision_2024}
\bibfield{author}{\bibinfo{person}{Patrick Grady}, \bibinfo{person}{Jeremy~A Collins}, \bibinfo{person}{Chengcheng Tang}, \bibinfo{person}{Christopher~D Twigg}, \bibinfo{person}{Kunal Aneja}, \bibinfo{person}{James Hays}, {and} \bibinfo{person}{Charles~C Kemp}.} \bibinfo{year}{2024}\natexlab{}.
\newblock \showarticletitle{{PressureVision++}: Estimating Fingertip Pressure from Diverse RGB Images}.
\newblock \bibinfo{journal}{\emph{IEEE/CVF Winter Conference on Applications of Computer Vision (WACV)}} (\bibinfo{year}{2024}).
\newblock


\bibitem[Grady et~al\mbox{.}(2022)]%
        {grady_pressurevision_2022}
\bibfield{author}{\bibinfo{person}{Patrick Grady}, \bibinfo{person}{Chengcheng Tang}, \bibinfo{person}{Samarth Brahmbhatt}, \bibinfo{person}{Christopher~D. Twigg}, \bibinfo{person}{Chengde Wan}, \bibinfo{person}{James Hays}, {and} \bibinfo{person}{Charles~C. Kemp}.} \bibinfo{year}{2022}\natexlab{}.
\newblock \showarticletitle{{PressureVision}: {Estimating} {Hand} {Pressure} from a {Single} {RGB} {Image}}. In \bibinfo{booktitle}{\emph{Computer {Vision} – {ECCV} 2022: 17th {European} {Conference}, {Tel} {Aviv}, {Israel}, {October} 23–27, 2022, {Proceedings}, {Part} {VI}}}. \bibinfo{publisher}{Springer-Verlag}, \bibinfo{address}{Berlin, Heidelberg}, \bibinfo{pages}{328--345}.
\newblock
\showISBNx{9783031200670}
\urldef\tempurl%
\url{https://doi.org/10.1007/978-3-031-20068-7_19}
\showDOI{\tempurl}


\bibitem[Gu et~al\mbox{.}(2019)]%
        {gu_accurate_2019}
\bibfield{author}{\bibinfo{person}{Yizheng Gu}, \bibinfo{person}{Chun Yu}, \bibinfo{person}{Zhipeng Li}, \bibinfo{person}{Weiqi Li}, \bibinfo{person}{Shuchang Xu}, \bibinfo{person}{Xiaoying Wei}, {and} \bibinfo{person}{Yuanchun Shi}.} \bibinfo{year}{2019}\natexlab{}.
\newblock \showarticletitle{Accurate and {Low}-{Latency} {Sensing} of {Touch} {Contact} on {Any} {Surface} with {Finger}-{Worn} {IMU} {Sensor}}. In \bibinfo{booktitle}{\emph{Proceedings of the 32nd {Annual} {ACM} {Symposium} on {User} {Interface} {Software} and {Technology}}} \emph{(\bibinfo{series}{{UIST} '19})}. \bibinfo{publisher}{Association for Computing Machinery}, \bibinfo{address}{New York, NY, USA}, \bibinfo{pages}{1059--1070}.
\newblock
\showISBNx{978-1-4503-6816-2}
\urldef\tempurl%
\url{https://doi.org/10.1145/3332165.3347947}
\showDOI{\tempurl}


\bibitem[Gu et~al\mbox{.}(2020)]%
        {gu_qwertyring_2020}
\bibfield{author}{\bibinfo{person}{Yizheng Gu}, \bibinfo{person}{Chun Yu}, \bibinfo{person}{Zhipeng Li}, \bibinfo{person}{Zhaoheng Li}, \bibinfo{person}{Xiaoying Wei}, {and} \bibinfo{person}{Yuanchun Shi}.} \bibinfo{year}{2020}\natexlab{}.
\newblock \showarticletitle{{QwertyRing}: {Text} {Entry} on {Physical} {Surfaces} {Using} a {Ring}}.
\newblock \bibinfo{journal}{\emph{Proc. ACM Interact. Mob. Wearable Ubiquitous Technol.}} \bibinfo{volume}{4}, \bibinfo{number}{4} (\bibinfo{date}{Dec.} \bibinfo{year}{2020}), \bibinfo{pages}{128:1--128:29}.
\newblock
\urldef\tempurl%
\url{https://doi.org/10.1145/3432204}
\showDOI{\tempurl}


\bibitem[Gustafson et~al\mbox{.}(2011)]%
        {gustafson_imaginary_2011}
\bibfield{author}{\bibinfo{person}{Sean Gustafson}, \bibinfo{person}{Christian Holz}, {and} \bibinfo{person}{Patrick Baudisch}.} \bibinfo{year}{2011}\natexlab{}.
\newblock \showarticletitle{Imaginary phone: learning imaginary interfaces by transferring spatial memory from a familiar device}. In \bibinfo{booktitle}{\emph{Proceedings of the 24th annual {ACM} symposium on {User} interface software and technology}}. \bibinfo{publisher}{ACM}, \bibinfo{address}{Santa Barbara California USA}, \bibinfo{pages}{283--292}.
\newblock
\showISBNx{978-1-4503-0716-1}
\urldef\tempurl%
\url{https://doi.org/10.1145/2047196.2047233}
\showDOI{\tempurl}


\bibitem[Harrison et~al\mbox{.}(2011)]%
        {harrison_omnitouch_2011}
\bibfield{author}{\bibinfo{person}{Chris Harrison}, \bibinfo{person}{Hrvoje Benko}, {and} \bibinfo{person}{Andrew~D. Wilson}.} \bibinfo{year}{2011}\natexlab{}.
\newblock \showarticletitle{{OmniTouch}: wearable multitouch interaction everywhere}. In \bibinfo{booktitle}{\emph{Proceedings of the 24th annual {ACM} symposium on {User} interface software and technology}} \emph{(\bibinfo{series}{{UIST} '11})}. \bibinfo{publisher}{Association for Computing Machinery}, \bibinfo{address}{New York, NY, USA}, \bibinfo{pages}{441--450}.
\newblock
\showISBNx{978-1-4503-0716-1}
\urldef\tempurl%
\url{https://doi.org/10.1145/2047196.2047255}
\showDOI{\tempurl}


\bibitem[Harrison and Hudson(2008)]%
        {scratchInput}
\bibfield{author}{\bibinfo{person}{Chris Harrison} {and} \bibinfo{person}{Scott~E. Hudson}.} \bibinfo{year}{2008}\natexlab{}.
\newblock \showarticletitle{Scratch input: creating large, inexpensive, unpowered and mobile finger input surfaces}. In \bibinfo{booktitle}{\emph{Proceedings of the 21st Annual ACM Symposium on User Interface Software and Technology}} (Monterey, CA, USA) \emph{(\bibinfo{series}{UIST '08})}. \bibinfo{publisher}{Association for Computing Machinery}, \bibinfo{address}{New York, NY, USA}, \bibinfo{pages}{205–208}.
\newblock
\showISBNx{9781595939753}
\urldef\tempurl%
\url{https://doi.org/10.1145/1449715.1449747}
\showDOI{\tempurl}


\bibitem[Harrison et~al\mbox{.}(2010)]%
        {harrison_skinput_2010}
\bibfield{author}{\bibinfo{person}{Chris Harrison}, \bibinfo{person}{Desney Tan}, {and} \bibinfo{person}{Dan Morris}.} \bibinfo{year}{2010}\natexlab{}.
\newblock \showarticletitle{Skinput: appropriating the body as an input surface}. In \bibinfo{booktitle}{\emph{Proceedings of the {SIGCHI} {Conference} on {Human} {Factors} in {Computing} {Systems}}} \emph{(\bibinfo{series}{{CHI} '10})}. \bibinfo{publisher}{Association for Computing Machinery}, \bibinfo{address}{New York, NY, USA}, \bibinfo{pages}{453--462}.
\newblock
\showISBNx{978-1-60558-929-9}
\urldef\tempurl%
\url{https://doi.org/10.1145/1753326.1753394}
\showDOI{\tempurl}


\bibitem[He et~al\mbox{.}(2025)]%
        {palmpadchi2025}
\bibfield{author}{\bibinfo{person}{Zhe He}, \bibinfo{person}{Xiangyang Wang}, \bibinfo{person}{Yuanchun Shi}, \bibinfo{person}{Chi Hsia}, \bibinfo{person}{Chen Liang}, {and} \bibinfo{person}{Chun Yu}.} \bibinfo{year}{2025}\natexlab{}.
\newblock \showarticletitle{Palmpad: Enabling Real-Time Index-to-Palm Touch Interaction with a Single RGB Camera}. In \bibinfo{booktitle}{\emph{Proceedings of the 2025 CHI Conference on Human Factors in Computing Systems}} \emph{(\bibinfo{series}{CHI '25})}. \bibinfo{publisher}{Association for Computing Machinery}, \bibinfo{address}{New York, NY, USA}, Article \bibinfo{articleno}{551}, \bibinfo{numpages}{16}~pages.
\newblock
\showISBNx{9798400713941}
\urldef\tempurl%
\url{https://doi.org/10.1145/3706598.3714130}
\showDOI{\tempurl}


\bibitem[Heaney(2024)]%
        {heaney_quest_2024}
\bibfield{author}{\bibinfo{person}{David Heaney}.} \bibinfo{year}{2024}\natexlab{}.
\newblock \bibinfo{title}{Quest {3S} {Has} {Better} {Low}-{Light} {Hand} {Tracking} {Than} {Quest} 3}.
\newblock
\newblock
\urldef\tempurl%
\url{https://www.uploadvr.com/quest3s-hand-tracking-better-than-quest-3/}
\showURL{%
\tempurl}


\bibitem[Hendrycks and Gimpel(2023)]%
        {hendrycks_gaussian_2023}
\bibfield{author}{\bibinfo{person}{Dan Hendrycks} {and} \bibinfo{person}{Kevin Gimpel}.} \bibinfo{year}{2023}\natexlab{}.
\newblock \bibinfo{title}{Gaussian {Error} {Linear} {Units} ({GELUs})}.
\newblock
\newblock
\urldef\tempurl%
\url{https://doi.org/10.48550/arXiv.1606.08415}
\showDOI{\tempurl}
\newblock
\shownote{arXiv:1606.08415 [cs]}.


\bibitem[Hu et~al\mbox{.}(2020)]%
        {hu_shadowsense_2020}
\bibfield{author}{\bibinfo{person}{Yuhan Hu}, \bibinfo{person}{Sara~Maria Bejarano}, {and} \bibinfo{person}{Guy Hoffman}.} \bibinfo{year}{2020}\natexlab{}.
\newblock \showarticletitle{{ShadowSense}: {Detecting} {Human} {Touch} in a {Social} {Robot} {Using} {Shadow} {Image} {Classification}}.
\newblock \bibinfo{journal}{\emph{Proc. ACM Interact. Mob. Wearable Ubiquitous Technol.}} \bibinfo{volume}{4}, \bibinfo{number}{4} (\bibinfo{date}{Dec.} \bibinfo{year}{2020}), \bibinfo{pages}{132:1--132:24}.
\newblock
\urldef\tempurl%
\url{https://doi.org/10.1145/3432202}
\showDOI{\tempurl}


\bibitem[Iacolina et~al\mbox{.}(2011)]%
        {iacolina_improving_2011}
\bibfield{author}{\bibinfo{person}{Samuel~A. Iacolina}, \bibinfo{person}{Alessandro Soro}, {and} \bibinfo{person}{Riccardo Scateni}.} \bibinfo{year}{2011}\natexlab{}.
\newblock \showarticletitle{Improving {FTIR} based multi-touch sensors with {IR} shadow tracking}. In \bibinfo{booktitle}{\emph{Proceedings of the 3rd {ACM} {SIGCHI} symposium on {Engineering} interactive computing systems}}. \bibinfo{publisher}{ACM}, \bibinfo{address}{Pisa Italy}, \bibinfo{pages}{241--246}.
\newblock
\showISBNx{978-1-4503-0670-6}
\urldef\tempurl%
\url{https://doi.org/10.1145/1996461.1996529}
\showDOI{\tempurl}


\bibitem[Iwai and Sato(2005)]%
        {iwai_heat_2005}
\bibfield{author}{\bibinfo{person}{Daisuke Iwai} {and} \bibinfo{person}{Kosuke Sato}.} \bibinfo{year}{2005}\natexlab{}.
\newblock \showarticletitle{Heat sensation in image creation with thermal vision}. In \bibinfo{booktitle}{\emph{Proceedings of the 2005 {ACM} {SIGCHI} {International} {Conference} on {Advances} in computer entertainment technology}} \emph{(\bibinfo{series}{{ACE} '05})}. \bibinfo{publisher}{Association for Computing Machinery}, \bibinfo{address}{New York, NY, USA}, \bibinfo{pages}{213--216}.
\newblock
\showISBNx{978-1-59593-110-8}
\urldef\tempurl%
\url{https://doi.org/10.1145/1178477.1178510}
\showDOI{\tempurl}


\bibitem[Kienzle and Hinckley(2014)]%
        {kienzle_lightring_2014}
\bibfield{author}{\bibinfo{person}{Wolf Kienzle} {and} \bibinfo{person}{Ken Hinckley}.} \bibinfo{year}{2014}\natexlab{}.
\newblock \showarticletitle{{LightRing}: always-available {2D} input on any surface}. In \bibinfo{booktitle}{\emph{Proceedings of the 27th annual {ACM} symposium on {User} interface software and technology}}. \bibinfo{publisher}{ACM}, \bibinfo{address}{Honolulu Hawaii USA}, \bibinfo{pages}{157--160}.
\newblock
\showISBNx{978-1-4503-3069-5}
\urldef\tempurl%
\url{https://doi.org/10.1145/2642918.2647376}
\showDOI{\tempurl}


\bibitem[Kienzle et~al\mbox{.}(2021)]%
        {kienzle_electroring_2021}
\bibfield{author}{\bibinfo{person}{Wolf Kienzle}, \bibinfo{person}{Eric Whitmire}, \bibinfo{person}{Chris Rittaler}, {and} \bibinfo{person}{Hrvoje Benko}.} \bibinfo{year}{2021}\natexlab{}.
\newblock \showarticletitle{{ElectroRing}: {Subtle} {Pinch} and {Touch} {Detection} with a {Ring}}. In \bibinfo{booktitle}{\emph{Proceedings of the 2021 {CHI} {Conference} on {Human} {Factors} in {Computing} {Systems}}} \emph{(\bibinfo{series}{{CHI} '21})}. \bibinfo{publisher}{Association for Computing Machinery}, \bibinfo{address}{New York, NY, USA}, \bibinfo{pages}{1--12}.
\newblock
\showISBNx{9781450380966}
\urldef\tempurl%
\url{https://doi.org/10.1145/3411764.3445094}
\showDOI{\tempurl}


\bibitem[Kim et~al\mbox{.}(2021)]%
        {atatouch}
\bibfield{author}{\bibinfo{person}{Daehwa Kim}, \bibinfo{person}{Keunwoo Park}, {and} \bibinfo{person}{Geehyuk Lee}.} \bibinfo{year}{2021}\natexlab{}.
\newblock \showarticletitle{AtaTouch: Robust Finger Pinch Detection for a VR Controller Using RF Return Loss}. In \bibinfo{booktitle}{\emph{Proceedings of the 2021 CHI Conference on Human Factors in Computing Systems}} (Yokohama, Japan) \emph{(\bibinfo{series}{CHI '21})}. \bibinfo{publisher}{Association for Computing Machinery}, \bibinfo{address}{New York, NY, USA}, Article \bibinfo{articleno}{11}, \bibinfo{numpages}{9}~pages.
\newblock
\showISBNx{9781450380966}
\urldef\tempurl%
\url{https://doi.org/10.1145/3411764.3445442}
\showDOI{\tempurl}


\bibitem[Kim et~al\mbox{.}(2024)]%
        {kim_soundscroll_2024}
\bibfield{author}{\bibinfo{person}{Daehwa Kim}, \bibinfo{person}{Eric Whitmire}, \bibinfo{person}{Roger Boldu}, \bibinfo{person}{Wolf Kienzle}, {and} \bibinfo{person}{Hrvoje Benko}.} \bibinfo{year}{2024}\natexlab{}.
\newblock \showarticletitle{{SoundScroll}: {Robust} {Finger} {Slide} {Detection} {Using} {Friction} {Sound} and {Wrist}-{Worn} {Microphones}}. In \bibinfo{booktitle}{\emph{Proceedings of the 2024 {ACM} {International} {Symposium} on {Wearable} {Computers}}} \emph{(\bibinfo{series}{{ISWC} '24})}. \bibinfo{publisher}{Association for Computing Machinery}, \bibinfo{address}{New York, NY, USA}, \bibinfo{pages}{63--70}.
\newblock
\showISBNx{979-8-4007-1059-9}
\urldef\tempurl%
\url{https://doi.org/10.1145/3675095.3676614}
\showDOI{\tempurl}


\bibitem[Krueger et~al\mbox{.}(1985)]%
        {krueger_videoplaceartificial_1985}
\bibfield{author}{\bibinfo{person}{Myron~W. Krueger}, \bibinfo{person}{Thomas Gionfriddo}, {and} \bibinfo{person}{Katrin Hinrichsen}.} \bibinfo{year}{1985}\natexlab{}.
\newblock \showarticletitle{{VIDEOPLACE}—an artificial reality}. In \bibinfo{booktitle}{\emph{Proceedings of the {SIGCHI} {Conference} on {Human} {Factors} in {Computing} {Systems}}} \emph{(\bibinfo{series}{{CHI} '85})}. \bibinfo{publisher}{Association for Computing Machinery}, \bibinfo{address}{New York, NY, USA}, \bibinfo{pages}{35--40}.
\newblock
\showISBNx{978-0-89791-149-8}
\urldef\tempurl%
\url{https://doi.org/10.1145/317456.317463}
\showDOI{\tempurl}


\bibitem[Kurz(2014)]%
        {kurz_thermal_2014}
\bibfield{author}{\bibinfo{person}{Daniel Kurz}.} \bibinfo{year}{2014}\natexlab{}.
\newblock \showarticletitle{Thermal touch: {Thermography}-enabled everywhere touch interfaces for mobile augmented reality applications}. In \bibinfo{booktitle}{\emph{2014 {IEEE} {International} {Symposium} on {Mixed} and {Augmented} {Reality} ({ISMAR})}}. \bibinfo{pages}{9--16}.
\newblock
\urldef\tempurl%
\url{https://doi.org/10.1109/ISMAR.2014.6948403}
\showDOI{\tempurl}


\bibitem[Laput and Harrison(2019)]%
        {SurfaceSight}
\bibfield{author}{\bibinfo{person}{Gierad Laput} {and} \bibinfo{person}{Chris Harrison}.} \bibinfo{year}{2019}\natexlab{}.
\newblock \showarticletitle{SurfaceSight: A New Spin on Touch, User, and Object Sensing for IoT Experiences}. In \bibinfo{booktitle}{\emph{Proceedings of the 2019 CHI Conference on Human Factors in Computing Systems}} (Glasgow, Scotland Uk) \emph{(\bibinfo{series}{CHI '19})}. \bibinfo{publisher}{Association for Computing Machinery}, \bibinfo{address}{New York, NY, USA}, \bibinfo{pages}{1–12}.
\newblock
\showISBNx{9781450359702}
\urldef\tempurl%
\url{https://doi.org/10.1145/3290605.3300559}
\showDOI{\tempurl}


\bibitem[Larson et~al\mbox{.}(2011)]%
        {larson_heatwave_2011}
\bibfield{author}{\bibinfo{person}{Eric Larson}, \bibinfo{person}{Gabe Cohn}, \bibinfo{person}{Sidhant Gupta}, \bibinfo{person}{Xiaofeng Ren}, \bibinfo{person}{Beverly Harrison}, \bibinfo{person}{Dieter Fox}, {and} \bibinfo{person}{Shwetak Patel}.} \bibinfo{year}{2011}\natexlab{}.
\newblock \showarticletitle{{HeatWave}: thermal imaging for surface user interaction}. In \bibinfo{booktitle}{\emph{Proceedings of the {SIGCHI} {Conference} on {Human} {Factors} in {Computing} {Systems}}} \emph{(\bibinfo{series}{{CHI} '11})}. \bibinfo{publisher}{Association for Computing Machinery}, \bibinfo{address}{New York, NY, USA}, \bibinfo{pages}{2565--2574}.
\newblock
\showISBNx{978-1-4503-0228-9}
\urldef\tempurl%
\url{https://doi.org/10.1145/1978942.1979317}
\showDOI{\tempurl}


\bibitem[Liang et~al\mbox{.}(2023)]%
        {liang_shadowtouch_2023}
\bibfield{author}{\bibinfo{person}{Chen Liang}, \bibinfo{person}{Xutong Wang}, \bibinfo{person}{Zisu Li}, \bibinfo{person}{Chi Hsia}, \bibinfo{person}{Mingming Fan}, \bibinfo{person}{Chun Yu}, {and} \bibinfo{person}{Yuanchun Shi}.} \bibinfo{year}{2023}\natexlab{}.
\newblock \showarticletitle{{ShadowTouch}: {Enabling} {Free}-{Form} {Touch}-{Based} {Hand}-to-{Surface} {Interaction} with {Wrist}-{Mounted} {Illuminant} by {Shadow} {Projection}}. In \bibinfo{booktitle}{\emph{Proceedings of the 36th {Annual} {ACM} {Symposium} on {User} {Interface} {Software} and {Technology}}} \emph{(\bibinfo{series}{{UIST} '23})}. \bibinfo{publisher}{Association for Computing Machinery}, \bibinfo{address}{New York, NY, USA}, \bibinfo{pages}{1--14}.
\newblock
\showISBNx{9798400701320}
\urldef\tempurl%
\url{https://doi.org/10.1145/3586183.3606785}
\showDOI{\tempurl}


\bibitem[Liang et~al\mbox{.}(2021)]%
        {liang_dualring_2021}
\bibfield{author}{\bibinfo{person}{Chen Liang}, \bibinfo{person}{Chun Yu}, \bibinfo{person}{Yue Qin}, \bibinfo{person}{Yuntao Wang}, {and} \bibinfo{person}{Yuanchun Shi}.} \bibinfo{year}{2021}\natexlab{}.
\newblock \showarticletitle{{DualRing}: {Enabling} {Subtle} and {Expressive} {Hand} {Interaction} with {Dual} {IMU} {Rings}}.
\newblock \bibinfo{journal}{\emph{Proceedings of the ACM on Interactive, Mobile, Wearable and Ubiquitous Technologies}} \bibinfo{volume}{5}, \bibinfo{number}{3} (\bibinfo{date}{Sept.} \bibinfo{year}{2021}), \bibinfo{pages}{1--27}.
\newblock
\showISSN{2474-9567}
\urldef\tempurl%
\url{https://doi.org/10.1145/3478114}
\showDOI{\tempurl}


\bibitem[Marshall et~al\mbox{.}(2009)]%
        {marshall_pressing_2009}
\bibfield{author}{\bibinfo{person}{Joe Marshall}, \bibinfo{person}{Tony Pridmore}, \bibinfo{person}{Mike Pound}, \bibinfo{person}{Steve Benford}, {and} \bibinfo{person}{Boriana Koleva}.} \bibinfo{year}{2009}\natexlab{}.
\newblock \showarticletitle{Pressing the {Flesh}: {Sensing} {Multiple} {Touch} and {Finger} {Pressure} on {Arbitrary} {Surfaces}}. In \bibinfo{booktitle}{\emph{Proceedings of the 6th {International} {Conference} on {Pervasive} {Computing}}} \emph{(\bibinfo{series}{Pervasive '08})}. \bibinfo{publisher}{Springer-Verlag}, \bibinfo{address}{Berlin, Heidelberg}, \bibinfo{pages}{38--55}.
\newblock
\showISBNx{978-3-540-79575-9}
\urldef\tempurl%
\url{https://doi.org/10.1007/978-3-540-79576-6_3}
\showDOI{\tempurl}


\bibitem[Masson et~al\mbox{.}(2017)]%
        {masson_whichfingers_2017}
\bibfield{author}{\bibinfo{person}{Damien Masson}, \bibinfo{person}{Alix Goguey}, \bibinfo{person}{Sylvain Malacria}, {and} \bibinfo{person}{Géry Casiez}.} \bibinfo{year}{2017}\natexlab{}.
\newblock \showarticletitle{{WhichFingers}: {Identifying} {Fingers} on {Touch} {Surfaces} and {Keyboards} using {Vibration} {Sensors}}. In \bibinfo{booktitle}{\emph{Proceedings of the 30th {Annual} {ACM} {Symposium} on {User} {Interface} {Software} and {Technology}}} \emph{(\bibinfo{series}{{UIST} '17})}. \bibinfo{publisher}{Association for Computing Machinery}, \bibinfo{address}{New York, NY, USA}, \bibinfo{pages}{41--48}.
\newblock
\showISBNx{978-1-4503-4981-9}
\urldef\tempurl%
\url{https://doi.org/10.1145/3126594.3126619}
\showDOI{\tempurl}


\bibitem[Matsubara et~al\mbox{.}(2017)]%
        {matsubara_touch_2017}
\bibfield{author}{\bibinfo{person}{Takashi Matsubara}, \bibinfo{person}{Naoki Mori}, \bibinfo{person}{Takehiro Niikura}, {and} \bibinfo{person}{Shun'ichi Tano}.} \bibinfo{year}{2017}\natexlab{}.
\newblock \showarticletitle{Touch detection method for non-display surface using multiple shadows of finger}. In \bibinfo{booktitle}{\emph{2017 {IEEE} 6th {Global} {Conference} on {Consumer} {Electronics} ({GCCE})}}. \bibinfo{pages}{1--5}.
\newblock
\urldef\tempurl%
\url{https://doi.org/10.1109/GCCE.2017.8229364}
\showDOI{\tempurl}


\bibitem[Meier et~al\mbox{.}(2021)]%
        {meier_tapid_2021}
\bibfield{author}{\bibinfo{person}{Manuel Meier}, \bibinfo{person}{Paul Streli}, \bibinfo{person}{Andreas Fender}, {and} \bibinfo{person}{Christian Holz}.} \bibinfo{year}{2021}\natexlab{}.
\newblock \showarticletitle{{TapID}: {Rapid} {Touch} {Interaction} in {Virtual} {Reality} using {Wearable} {Sensing}}. In \bibinfo{booktitle}{\emph{2021 {IEEE} {Virtual} {Reality} and {3D} {User} {Interfaces} ({VR})}}. \bibinfo{pages}{519--528}.
\newblock
\urldef\tempurl%
\url{https://doi.org/10.1109/VR50410.2021.00076}
\showDOI{\tempurl}
\newblock
\shownote{ISSN: 2642-5254}.


\bibitem[Mollyn and Harrison(2024)]%
        {mollyn_egotouch_2024}
\bibfield{author}{\bibinfo{person}{Vimal Mollyn} {and} \bibinfo{person}{Chris Harrison}.} \bibinfo{year}{2024}\natexlab{}.
\newblock \showarticletitle{{EgoTouch}: {On}-{Body} {Touch} {Input} {Using} {AR}/{VR} {Headset} {Cameras}}. In \bibinfo{booktitle}{\emph{Proceedings of the 37th {Annual} {ACM} {Symposium} on {User} {Interface} {Software} and {Technology}}} \emph{(\bibinfo{series}{{UIST} '24})}. \bibinfo{publisher}{Association for Computing Machinery}, \bibinfo{address}{New York, NY, USA}, \bibinfo{pages}{1--11}.
\newblock
\showISBNx{9798400706288}
\urldef\tempurl%
\url{https://doi.org/10.1145/3654777.3676455}
\showDOI{\tempurl}


\bibitem[Mujibiya et~al\mbox{.}(2013)]%
        {mujibiya_sound_2013}
\bibfield{author}{\bibinfo{person}{Adiyan Mujibiya}, \bibinfo{person}{Xiang Cao}, \bibinfo{person}{Desney~S. Tan}, \bibinfo{person}{Dan Morris}, \bibinfo{person}{Shwetak~N. Patel}, {and} \bibinfo{person}{Jun Rekimoto}.} \bibinfo{year}{2013}\natexlab{}.
\newblock \showarticletitle{The sound of touch: on-body touch and gesture sensing based on transdermal ultrasound propagation}. In \bibinfo{booktitle}{\emph{Proceedings of the 2013 {ACM} international conference on {Interactive} tabletops and surfaces}} \emph{(\bibinfo{series}{{ITS} '13})}. \bibinfo{publisher}{Association for Computing Machinery}, \bibinfo{address}{New York, NY, USA}, \bibinfo{pages}{189--198}.
\newblock
\showISBNx{978-1-4503-2271-3}
\urldef\tempurl%
\url{https://doi.org/10.1145/2512349.2512821}
\showDOI{\tempurl}


\bibitem[Niikura et~al\mbox{.}(2016)]%
        {niikura_touch_2016}
\bibfield{author}{\bibinfo{person}{Takehiro Niikura}, \bibinfo{person}{Takashi Matsubara}, {and} \bibinfo{person}{Naoki Mori}.} \bibinfo{year}{2016}\natexlab{}.
\newblock \showarticletitle{Touch Detection System for Various Surfaces Using Shadow of Finger}. In \bibinfo{booktitle}{\emph{Proceedings of the 2016 ACM International Conference on Interactive Surfaces and Spaces}} (Niagara Falls, Ontario, Canada) \emph{(\bibinfo{series}{ISS '16})}. \bibinfo{publisher}{Association for Computing Machinery}, \bibinfo{address}{New York, NY, USA}, \bibinfo{pages}{337–342}.
\newblock
\showISBNx{9781450342483}
\urldef\tempurl%
\url{https://doi.org/10.1145/2992154.2996777}
\showDOI{\tempurl}


\bibitem[Oh et~al\mbox{.}(2017)]%
        {oh_anywheretouch_2017}
\bibfield{author}{\bibinfo{person}{Ju~Young Oh}, \bibinfo{person}{Jun Lee}, \bibinfo{person}{Joong~Ho Lee}, {and} \bibinfo{person}{Ji~Hyung Park}.} \bibinfo{year}{2017}\natexlab{}.
\newblock \showarticletitle{{AnywhereTouch}: {Finger} {Tracking} {Method} on {Arbitrary} {Surface} {Using} {Nailed}-{Mounted} {IMU} for {Mobile} {HMD}}. In \bibinfo{booktitle}{\emph{{HCI} {International} 2017 – {Posters}' {Extended} {Abstracts}}}, \bibfield{editor}{\bibinfo{person}{Constantine Stephanidis}} (Ed.). \bibinfo{publisher}{Springer International Publishing}, \bibinfo{address}{Cham}, \bibinfo{pages}{185--191}.
\newblock
\showISBNx{978-3-319-58750-9}
\urldef\tempurl%
\url{https://doi.org/10.1007/978-3-319-58750-9_26}
\showDOI{\tempurl}


\bibitem[Oh et~al\mbox{.}(2020)]%
        {oh_fingertouch_2020}
\bibfield{author}{\bibinfo{person}{Ju~Young Oh}, \bibinfo{person}{Ji-Hyung Park}, {and} \bibinfo{person}{Jung-Min Park}.} \bibinfo{year}{2020}\natexlab{}.
\newblock \showarticletitle{{FingerTouch}: {Touch} {Interaction} {Using} a {Fingernail}-{Mounted} {Sensor} on a {Head}-{Mounted} {Display} for {Augmented} {Reality}}.
\newblock \bibinfo{journal}{\emph{IEEE Access}}  \bibinfo{volume}{8} (\bibinfo{year}{2020}), \bibinfo{pages}{101192--101208}.
\newblock
\showISSN{2169-3536}
\urldef\tempurl%
\url{https://doi.org/10.1109/ACCESS.2020.2997972}
\showDOI{\tempurl}
\newblock
\shownote{Conference Name: IEEE Access}.


\bibitem[Ono et~al\mbox{.}(2013)]%
        {ono_touch_2013}
\bibfield{author}{\bibinfo{person}{Makoto Ono}, \bibinfo{person}{Buntarou Shizuki}, {and} \bibinfo{person}{Jiro Tanaka}.} \bibinfo{year}{2013}\natexlab{}.
\newblock \showarticletitle{Touch \& activate: adding interactivity to existing objects using active acoustic sensing}. In \bibinfo{booktitle}{\emph{Proceedings of the 26th annual {ACM} symposium on {User} interface software and technology}}. \bibinfo{publisher}{ACM}, \bibinfo{address}{St. Andrews Scotland, United Kingdom}, \bibinfo{pages}{31--40}.
\newblock
\showISBNx{978-1-4503-2268-3}
\urldef\tempurl%
\url{https://doi.org/10.1145/2501988.2501989}
\showDOI{\tempurl}


\bibitem[{Pak-Kiu Chung} et~al\mbox{.}(2008)]%
        {pak-kiu_chung_mirrortrack_2008}
\bibfield{author}{\bibinfo{person}{{Pak-Kiu Chung}}, \bibinfo{person}{{Bing Fang}}, {and} \bibinfo{person}{F. Quek}.} \bibinfo{year}{2008}\natexlab{}.
\newblock \showarticletitle{{MirrorTrack} - a vision based multi-touch system for glossy display surfaces}. In \bibinfo{booktitle}{\emph{5th {International} {Conference} on {Visual} {Information} {Engineering} ({VIE} 2008)}}. \bibinfo{publisher}{IEE}, \bibinfo{address}{Xi'an, China}, \bibinfo{pages}{571--576}.
\newblock
\showISBNx{978-0-86341-914-0}
\urldef\tempurl%
\url{https://doi.org/10.1049/cp:20080379}
\showDOI{\tempurl}


\bibitem[Pei et~al\mbox{.}(2022)]%
        {pei_forcesight_2022}
\bibfield{author}{\bibinfo{person}{Siyou Pei}, \bibinfo{person}{Pradyumna Chari}, \bibinfo{person}{Xue Wang}, \bibinfo{person}{Xiaoying Yang}, \bibinfo{person}{Achuta Kadambi}, {and} \bibinfo{person}{Yang Zhang}.} \bibinfo{year}{2022}\natexlab{}.
\newblock \showarticletitle{ForceSight: Non-Contact Force Sensing with Laser Speckle Imaging}. In \bibinfo{booktitle}{\emph{Proceedings of the 35th Annual ACM Symposium on User Interface Software and Technology}} (Bend, OR, USA) \emph{(\bibinfo{series}{UIST '22})}. \bibinfo{publisher}{Association for Computing Machinery}, \bibinfo{address}{New York, NY, USA}, Article \bibinfo{articleno}{25}, \bibinfo{numpages}{11}~pages.
\newblock
\showISBNx{9781450393201}
\urldef\tempurl%
\url{https://doi.org/10.1145/3526113.3545622}
\showDOI{\tempurl}


\bibitem[Pham et~al\mbox{.}(2005)]%
        {pham_tangible_2002}
\bibfield{author}{\bibinfo{person}{DT Pham}, \bibinfo{person}{M Al-Kutubi}, \bibinfo{person}{Z Ji}, \bibinfo{person}{M Yang}, \bibinfo{person}{Z Wang}, {and} \bibinfo{person}{S Catheline}.} \bibinfo{year}{2005}\natexlab{}.
\newblock \showarticletitle{Tangible acoustic interface approaches}. In \bibinfo{booktitle}{\emph{Proceedings of IPROMS 2005 Virtual Conference}}. Citeseer, \bibinfo{pages}{497--502}.
\newblock


\bibitem[Posner et~al\mbox{.}(2012)]%
        {posner_single_2012}
\bibfield{author}{\bibinfo{person}{Erez Posner}, \bibinfo{person}{Nick Starzicki}, {and} \bibinfo{person}{Eyal Katz}.} \bibinfo{year}{2012}\natexlab{}.
\newblock \showarticletitle{A single camera based floating virtual keyboard with improved touch detection}. In \bibinfo{booktitle}{\emph{2012 {IEEE} 27th {Convention} of {Electrical} and {Electronics} {Engineers} in {Israel}}}. \bibinfo{pages}{1--5}.
\newblock
\urldef\tempurl%
\url{https://doi.org/10.1109/EEEI.2012.6377072}
\showDOI{\tempurl}


\bibitem[Richardson et~al\mbox{.}(2020)]%
        {richardson_decoding_2020}
\bibfield{author}{\bibinfo{person}{Mark Richardson}, \bibinfo{person}{Matt Durasoff}, {and} \bibinfo{person}{Robert Wang}.} \bibinfo{year}{2020}\natexlab{}.
\newblock \showarticletitle{Decoding {Surface} {Touch} {Typing} from {Hand}-{Tracking}}. In \bibinfo{booktitle}{\emph{Proceedings of the 33rd {Annual} {ACM} {Symposium} on {User} {Interface} {Software} and {Technology}}}. \bibinfo{publisher}{ACM}, \bibinfo{address}{Virtual Event USA}, \bibinfo{pages}{686--696}.
\newblock
\showISBNx{978-1-4503-7514-6}
\urldef\tempurl%
\url{https://doi.org/10.1145/3379337.3415816}
\showDOI{\tempurl}


\bibitem[Shen et~al\mbox{.}(2021)]%
        {shen_farout_2021}
\bibfield{author}{\bibinfo{person}{Vivian Shen}, \bibinfo{person}{James Spann}, {and} \bibinfo{person}{Chris Harrison}.} \bibinfo{year}{2021}\natexlab{}.
\newblock \showarticletitle{{FarOut} {Touch}: {Extending} the {Range} of ad hoc {Touch} {Sensing} with {Depth} {Cameras}}. In \bibinfo{booktitle}{\emph{Symposium on {Spatial} {User} {Interaction}}}. \bibinfo{publisher}{ACM}, \bibinfo{address}{Virtual Event USA}, \bibinfo{pages}{1--12}.
\newblock
\showISBNx{978-1-4503-9091-0}
\urldef\tempurl%
\url{https://doi.org/10.1145/3485279.3485281}
\showDOI{\tempurl}


\bibitem[Shen et~al\mbox{.}(2024)]%
        {shen_mousering_2024}
\bibfield{author}{\bibinfo{person}{Xiyuan Shen}, \bibinfo{person}{Chun Yu}, \bibinfo{person}{Xutong Wang}, \bibinfo{person}{Chen Liang}, \bibinfo{person}{Haozhan Chen}, {and} \bibinfo{person}{Yuanchun Shi}.} \bibinfo{year}{2024}\natexlab{}.
\newblock \showarticletitle{{MouseRing}: {Always}-available {Touchpad} {Interaction} with {IMU} {Rings}}. In \bibinfo{booktitle}{\emph{Proceedings of the 2024 {CHI} {Conference} on {Human} {Factors} in {Computing} {Systems}}} \emph{(\bibinfo{series}{{CHI} '24})}. \bibinfo{publisher}{Association for Computing Machinery}, \bibinfo{address}{New York, NY, USA}, \bibinfo{pages}{1--19}.
\newblock
\showISBNx{9798400703300}
\urldef\tempurl%
\url{https://doi.org/10.1145/3613904.3642225}
\showDOI{\tempurl}


\bibitem[Shi et~al\mbox{.}(2020a)]%
        {shi_versatouch_2020}
\bibfield{author}{\bibinfo{person}{Yilei Shi}, \bibinfo{person}{Haimo Zhang}, \bibinfo{person}{Jiashuo Cao}, {and} \bibinfo{person}{Suranga Nanayakkara}.} \bibinfo{year}{2020}\natexlab{a}.
\newblock \showarticletitle{{VersaTouch}: {A} {Versatile} {Plug}-and-{Play} {System} that {Enables} {Touch} {Interactions} on {Everyday} {Passive} {Surfaces}}. In \bibinfo{booktitle}{\emph{Proceedings of the {Augmented} {Humans} {International} {Conference}}} \emph{(\bibinfo{series}{{AHs} '20})}. \bibinfo{publisher}{Association for Computing Machinery}, \bibinfo{address}{New York, NY, USA}, \bibinfo{pages}{1--12}.
\newblock
\showISBNx{978-1-4503-7603-7}
\urldef\tempurl%
\url{https://doi.org/10.1145/3384657.3384778}
\showDOI{\tempurl}


\bibitem[Shi et~al\mbox{.}(2020b)]%
        {shi_ready_2020}
\bibfield{author}{\bibinfo{person}{Yilei Shi}, \bibinfo{person}{Haimo Zhang}, \bibinfo{person}{Kaixing Zhao}, \bibinfo{person}{Jiashuo Cao}, \bibinfo{person}{Mengmeng Sun}, {and} \bibinfo{person}{Suranga Nanayakkara}.} \bibinfo{year}{2020}\natexlab{b}.
\newblock \showarticletitle{Ready, {Steady}, {Touch}!: {Sensing} {Physical} {Contact} with a {Finger}-{Mounted} {IMU}}.
\newblock \bibinfo{journal}{\emph{Proceedings of the ACM on Interactive, Mobile, Wearable and Ubiquitous Technologies}} \bibinfo{volume}{4}, \bibinfo{number}{2} (\bibinfo{date}{June} \bibinfo{year}{2020}), \bibinfo{pages}{1--25}.
\newblock
\showISSN{2474-9567}
\urldef\tempurl%
\url{https://doi.org/10.1145/3397309}
\showDOI{\tempurl}


\bibitem[Shoemaker et~al\mbox{.}(2007)]%
        {shoemaker_shadow_2007}
\bibfield{author}{\bibinfo{person}{Garth Shoemaker}, \bibinfo{person}{Anthony Tang}, {and} \bibinfo{person}{Kellogg~S. Booth}.} \bibinfo{year}{2007}\natexlab{}.
\newblock \showarticletitle{Shadow reaching: a new perspective on interaction for large displays}.
\newblock \bibinfo{journal}{\emph{Proceedings of the 20th annual ACM symposium on User interface software and technology}} (\bibinfo{date}{Oct.} \bibinfo{year}{2007}), \bibinfo{pages}{53--56}.
\newblock
\urldef\tempurl%
\url{https://doi.org/10.1145/1294211.1294221}
\showDOI{\tempurl}
\newblock
\shownote{Conference Name: UIST07: The 20th Annual ACM Symposium on User Interface Software and Technology ISBN: 9781595936790 Place: Newport Rhode Island USA Publisher: ACM}.


\bibitem[Streli et~al\mbox{.}(2023)]%
        {streli_structured_2023}
\bibfield{author}{\bibinfo{person}{Paul Streli}, \bibinfo{person}{Jiaxi Jiang}, \bibinfo{person}{Juliete Rossie}, {and} \bibinfo{person}{Christian Holz}.} \bibinfo{year}{2023}\natexlab{}.
\newblock \showarticletitle{Structured {Light} {Speckle}: {Joint} {Ego}-{Centric} {Depth} {Estimation} and {Low}-{Latency} {Contact} {Detection} via {Remote} {Vibrometry}}. In \bibinfo{booktitle}{\emph{Proceedings of the 36th {Annual} {ACM} {Symposium} on {User} {Interface} {Software} and {Technology}}}. \bibinfo{publisher}{ACM}, \bibinfo{address}{San Francisco CA USA}, \bibinfo{pages}{1--12}.
\newblock
\showISBNx{9798400701320}
\urldef\tempurl%
\url{https://doi.org/10.1145/3586183.3606749}
\showDOI{\tempurl}


\bibitem[Streli et~al\mbox{.}(2024)]%
        {streli_touchinsight_2024}
\bibfield{author}{\bibinfo{person}{Paul Streli}, \bibinfo{person}{Mark Richardson}, \bibinfo{person}{Fadi Botros}, \bibinfo{person}{Shugao Ma}, \bibinfo{person}{Robert Wang}, {and} \bibinfo{person}{Christian Holz}.} \bibinfo{year}{2024}\natexlab{}.
\newblock \showarticletitle{{TouchInsight}: {Uncertainty}-aware {Rapid} {Touch} and {Text} {Input} for {Mixed} {Reality} from {Egocentric} {Vision}}. In \bibinfo{booktitle}{\emph{Proceedings of the 37th {Annual} {ACM} {Symposium} on {User} {Interface} {Software} and {Technology}}}. \bibinfo{publisher}{ACM}, \bibinfo{address}{Pittsburgh PA USA}, \bibinfo{pages}{1--16}.
\newblock
\showISBNx{9798400706288}
\urldef\tempurl%
\url{https://doi.org/10.1145/3654777.3676330}
\showDOI{\tempurl}


\bibitem[Strickon and Paradiso(1998)]%
        {strickon_tracking_1998}
\bibfield{author}{\bibinfo{person}{Joshua Strickon} {and} \bibinfo{person}{Joseph Paradiso}.} \bibinfo{year}{1998}\natexlab{}.
\newblock \showarticletitle{Tracking hands above large interactive surfaces with a low-cost scanning laser rangefinder}. In \bibinfo{booktitle}{\emph{{CHI} 98 {Conference} {Summary} on {Human} {Factors} in {Computing} {Systems}}} \emph{(\bibinfo{series}{{CHI} '98})}. \bibinfo{publisher}{Association for Computing Machinery}, \bibinfo{address}{New York, NY, USA}, \bibinfo{pages}{231--232}.
\newblock
\showISBNx{978-1-58113-028-7}
\urldef\tempurl%
\url{https://doi.org/10.1145/286498.286719}
\showDOI{\tempurl}


\bibitem[Sugita et~al\mbox{.}(2008)]%
        {sugita_touch_2008}
\bibfield{author}{\bibinfo{person}{Naoki Sugita}, \bibinfo{person}{Daisuke Iwai}, {and} \bibinfo{person}{Kosuke Sato}.} \bibinfo{year}{2008}\natexlab{}.
\newblock \showarticletitle{Touch sensing by image analysis of fingernail}. In \bibinfo{booktitle}{\emph{2008 {SICE} {Annual} {Conference}}}. \bibinfo{pages}{1520--1525}.
\newblock
\urldef\tempurl%
\url{https://doi.org/10.1109/SICE.2008.4654901}
\showDOI{\tempurl}


\bibitem[Thomas(2013)]%
        {thomas_camera_2013}
\bibfield{author}{\bibinfo{person}{Joseph Thomas}.} \bibinfo{year}{2013}\natexlab{}.
\newblock \showarticletitle{A Camera Based Virtual Keyboard with Touch Detection by Shadow Analysis}.
\newblock  (\bibinfo{year}{2013}).
\newblock
\urldef\tempurl%
\url{jsthomas.github.io/docs/vkeyboard/vkeyboard.pdf}
\showURL{%
\tempurl}


\bibitem[Vasu et~al\mbox{.}(2023)]%
        {vasu_fastvit_2023}
\bibfield{author}{\bibinfo{person}{Pavan Kumar~Anasosalu Vasu}, \bibinfo{person}{James Gabriel}, \bibinfo{person}{Jeff Zhu}, \bibinfo{person}{Oncel Tuzel}, {and} \bibinfo{person}{Anurag Ranjan}.} \bibinfo{year}{2023}\natexlab{}.
\newblock \showarticletitle{FastViT: A Fast Hybrid Vision Transformer using Structural Reparameterization}. In \bibinfo{booktitle}{\emph{Proceedings of the IEEE/CVF International Conference on Computer Vision}}.
\newblock


\bibitem[Vishal and Lawrence(2017)]%
        {vishal_paper_2017}
\bibfield{author}{\bibinfo{person}{Boga Vishal} {and} \bibinfo{person}{K~Deepak Lawrence}.} \bibinfo{year}{2017}\natexlab{}.
\newblock \showarticletitle{Paper piano — {Shadow} analysis based touch interaction}. In \bibinfo{booktitle}{\emph{2017 2nd {International} {Conference} on {Man} and {Machine} {Interfacing} ({MAMI})}}. \bibinfo{pages}{1--6}.
\newblock
\urldef\tempurl%
\url{https://doi.org/10.1109/MAMI.2017.8307890}
\showDOI{\tempurl}


\bibitem[Vision(2023)]%
        {noauthor_ht-sua33gm-t1v-c_nodate}
\bibfield{author}{\bibinfo{person}{Huateng Vision}.} \bibinfo{year}{2023}\natexlab{}.
\newblock \bibinfo{title}{{HT}-{SUA33GM}-{T1V}-{C} : {Huateng} {Vision}}.
\newblock
\newblock
\urldef\tempurl%
\url{https://huatengvision.com/product/195/}
\showURL{%
\tempurl}


\bibitem[Waghmare et~al\mbox{.}(2023)]%
        {waghmare_z-ring_2023}
\bibfield{author}{\bibinfo{person}{Anandghan Waghmare}, \bibinfo{person}{Youssef Ben~Taleb}, \bibinfo{person}{Ishan Chatterjee}, \bibinfo{person}{Arjun Narendra}, {and} \bibinfo{person}{Shwetak Patel}.} \bibinfo{year}{2023}\natexlab{}.
\newblock \showarticletitle{Z-{Ring}: {Single}-{Point} {Bio}-{Impedance} {Sensing} for {Gesture}, {Touch}, {Object} and {User} {Recognition}}. In \bibinfo{booktitle}{\emph{Proceedings of the 2023 {CHI} {Conference} on {Human} {Factors} in {Computing} {Systems}}}. \bibinfo{publisher}{ACM}, \bibinfo{address}{Hamburg Germany}, \bibinfo{pages}{1--18}.
\newblock
\showISBNx{978-1-4503-9421-5}
\urldef\tempurl%
\url{https://doi.org/10.1145/3544548.3581422}
\showDOI{\tempurl}


\bibitem[Wikipedia(2025)]%
        {noauthor_lux_2025}
\bibfield{author}{\bibinfo{person}{Wikipedia}.} \bibinfo{year}{2025}\natexlab{}.
\newblock \bibinfo{title}{Lux}.
\newblock
\newblock
\urldef\tempurl%
\url{https://en.wikipedia.org/w/index.php?title=Lux&oldid=1284696029}
\showURL{%
\tempurl}
\newblock
\shownote{Page Version ID: 1284696029}.


\bibitem[Wilson(2005)]%
        {wilson_playanywhere_2005}
\bibfield{author}{\bibinfo{person}{Andrew~D. Wilson}.} \bibinfo{year}{2005}\natexlab{}.
\newblock \showarticletitle{{PlayAnywhere}: a compact interactive tabletop projection-vision system}. In \bibinfo{booktitle}{\emph{Proceedings of the 18th annual {ACM} symposium on {User} interface software and technology}}. \bibinfo{publisher}{ACM}, \bibinfo{address}{Seattle WA USA}, \bibinfo{pages}{83--92}.
\newblock
\showISBNx{9781595932716}
\urldef\tempurl%
\url{https://doi.org/10.1145/1095034.1095047}
\showDOI{\tempurl}


\bibitem[Wilson(2010)]%
        {wilson_using_2010}
\bibfield{author}{\bibinfo{person}{Andrew~D. Wilson}.} \bibinfo{year}{2010}\natexlab{}.
\newblock \showarticletitle{Using a depth camera as a touch sensor}. In \bibinfo{booktitle}{\emph{{ACM} {International} {Conference} on {Interactive} {Tabletops} and {Surfaces}}}. \bibinfo{publisher}{ACM}, \bibinfo{address}{Saarbrücken Germany}, \bibinfo{pages}{69--72}.
\newblock
\showISBNx{978-1-4503-0399-6}
\urldef\tempurl%
\url{https://doi.org/10.1145/1936652.1936665}
\showDOI{\tempurl}


\bibitem[Xia et~al\mbox{.}(2025)]%
        {xia_halotouch_2025}
\bibfield{author}{\bibinfo{person}{Ziyi Xia}, \bibinfo{person}{Xincheng Huang}, \bibinfo{person}{Sidney~S Fels}, {and} \bibinfo{person}{Robert Xiao}.} \bibinfo{year}{2025}\natexlab{}.
\newblock \showarticletitle{HaloTouch: Using IR Multi-Path Interference to Support Touch Interactions with General Surfaces}. In \bibinfo{booktitle}{\emph{Proceedings of the 2025 CHI Conference on Human Factors in Computing Systems}} \emph{(\bibinfo{series}{CHI '25})}. \bibinfo{publisher}{Association for Computing Machinery}, \bibinfo{address}{New York, NY, USA}, Article \bibinfo{articleno}{548}, \bibinfo{numpages}{17}~pages.
\newblock
\showISBNx{9798400713941}
\urldef\tempurl%
\url{https://doi.org/10.1145/3706598.3714179}
\showDOI{\tempurl}


\bibitem[Xiao et~al\mbox{.}(2016)]%
        {xiao_direct_2016}
\bibfield{author}{\bibinfo{person}{Robert Xiao}, \bibinfo{person}{Scott Hudson}, {and} \bibinfo{person}{Chris Harrison}.} \bibinfo{year}{2016}\natexlab{}.
\newblock \showarticletitle{{DIRECT}: {Making} {Touch} {Tracking} on {Ordinary} {Surfaces} {Practical} with {Hybrid} {Depth}-{Infrared} {Sensing}}. In \bibinfo{booktitle}{\emph{Proceedings of the 2016 {ACM} {International} {Conference} on {Interactive} {Surfaces} and {Spaces}}}. \bibinfo{publisher}{ACM}, \bibinfo{address}{Niagara Falls Ontario Canada}, \bibinfo{pages}{85--94}.
\newblock
\showISBNx{978-1-4503-4248-3}
\urldef\tempurl%
\url{https://doi.org/10.1145/2992154.2992173}
\showDOI{\tempurl}


\bibitem[Xiao et~al\mbox{.}(2017)]%
        {xiao_desktop_2017}
\bibfield{author}{\bibinfo{person}{Robert Xiao}, \bibinfo{person}{Scott Hudson}, {and} \bibinfo{person}{Chris Harrison}.} \bibinfo{year}{2017}\natexlab{}.
\newblock \showarticletitle{Supporting Responsive Cohabitation Between Virtual Interfaces and Physical Objects on Everyday Surfaces}.
\newblock \bibinfo{journal}{\emph{Proc. ACM Hum.-Comput. Interact.}} \bibinfo{volume}{1}, \bibinfo{number}{EICS}, Article \bibinfo{articleno}{12} (\bibinfo{date}{June} \bibinfo{year}{2017}), \bibinfo{numpages}{17}~pages.
\newblock
\urldef\tempurl%
\url{https://doi.org/10.1145/3095814}
\showDOI{\tempurl}


\bibitem[Xiao et~al\mbox{.}(2018)]%
        {xiao_mrtouch_2018}
\bibfield{author}{\bibinfo{person}{Robert Xiao}, \bibinfo{person}{Julia Schwarz}, \bibinfo{person}{Nick Throm}, \bibinfo{person}{Andrew~D. Wilson}, {and} \bibinfo{person}{Hrvoje Benko}.} \bibinfo{year}{2018}\natexlab{}.
\newblock \showarticletitle{{MRTouch}: {Adding} {Touch} {Input} to {Head}-{Mounted} {Mixed} {Reality}}.
\newblock \bibinfo{journal}{\emph{IEEE Transactions on Visualization and Computer Graphics}} \bibinfo{volume}{24}, \bibinfo{number}{4} (\bibinfo{date}{April} \bibinfo{year}{2018}), \bibinfo{pages}{1653--1660}.
\newblock
\showISSN{1941-0506}
\urldef\tempurl%
\url{https://doi.org/10.1109/TVCG.2018.2794222}
\showDOI{\tempurl}


\bibitem[Yoo et~al\mbox{.}(2016)]%
        {yoo_symmetrisense_2016}
\bibfield{author}{\bibinfo{person}{Chungkuk Yoo}, \bibinfo{person}{Inseok Hwang}, \bibinfo{person}{Eric Rozner}, \bibinfo{person}{Yu Gu}, {and} \bibinfo{person}{Robert~F. Dickerson}.} \bibinfo{year}{2016}\natexlab{}.
\newblock \showarticletitle{{SymmetriSense}: {Enabling} {Near}-{Surface} {Interactivity} on {Glossy} {Surfaces} using a {Single} {Commodity} {Smartphone}}. In \bibinfo{booktitle}{\emph{Proceedings of the 2016 {CHI} {Conference} on {Human} {Factors} in {Computing} {Systems}}} \emph{(\bibinfo{series}{{CHI} '16})}. \bibinfo{publisher}{Association for Computing Machinery}, \bibinfo{address}{New York, NY, USA}, \bibinfo{pages}{5126--5137}.
\newblock
\showISBNx{978-1-4503-3362-7}
\urldef\tempurl%
\url{https://doi.org/10.1145/2858036.2858286}
\showDOI{\tempurl}


\bibitem[Zhang et~al\mbox{.}(2017)]%
        {zhang_soundtrak_2017}
\bibfield{author}{\bibinfo{person}{Cheng Zhang}, \bibinfo{person}{Qiuyue Xue}, \bibinfo{person}{Anandghan Waghmare}, \bibinfo{person}{Sumeet Jain}, \bibinfo{person}{Yiming Pu}, \bibinfo{person}{Sinan Hersek}, \bibinfo{person}{Kent Lyons}, \bibinfo{person}{Kenneth~A. Cunefare}, \bibinfo{person}{Omer~T. Inan}, {and} \bibinfo{person}{Gregory~D. Abowd}.} \bibinfo{year}{2017}\natexlab{}.
\newblock \showarticletitle{{SoundTrak}: {Continuous} {3D} {Tracking} of a {Finger} {Using} {Active} {Acoustics}}.
\newblock \bibinfo{journal}{\emph{Proceedings of the ACM on Interactive, Mobile, Wearable and Ubiquitous Technologies}} \bibinfo{volume}{1}, \bibinfo{number}{2} (\bibinfo{date}{June} \bibinfo{year}{2017}), \bibinfo{pages}{1--25}.
\newblock
\showISSN{2474-9567}
\urldef\tempurl%
\url{https://doi.org/10.1145/3090095}
\showDOI{\tempurl}


\bibitem[Zhang et~al\mbox{.}(2020)]%
        {zhang_mediapipe_2020}
\bibfield{author}{\bibinfo{person}{Fan Zhang}, \bibinfo{person}{Valentin Bazarevsky}, \bibinfo{person}{Andrey Vakunov}, \bibinfo{person}{Andrei Tkachenka}, \bibinfo{person}{George Sung}, \bibinfo{person}{Chuo-Ling Chang}, {and} \bibinfo{person}{Matthias Grundmann}.} \bibinfo{year}{2020}\natexlab{}.
\newblock \bibinfo{title}{{MediaPipe} {Hands}: {On}-device {Real}-time {Hand} {Tracking}}.
\newblock
\newblock
\urldef\tempurl%
\url{https://doi.org/10.48550/arXiv.2006.10214}
\showDOI{\tempurl}
\newblock
\shownote{arXiv:2006.10214 [cs]}.


\bibitem[Zhang et~al\mbox{.}(2025)]%
        {iclight2025}
\bibfield{author}{\bibinfo{person}{Lvmin Zhang}, \bibinfo{person}{Anyi Rao}, {and} \bibinfo{person}{Maneesh Agrawala}.} \bibinfo{year}{2025}\natexlab{}.
\newblock \showarticletitle{Scaling In-the-Wild Training for Diffusion-based Illumination Harmonization and Editing by Imposing Consistent Light Transport}. In \bibinfo{booktitle}{\emph{The Thirteenth International Conference on Learning Representations}}.
\newblock
\urldef\tempurl%
\url{https://openreview.net/forum?id=u1cQYxRI1H}
\showURL{%
\tempurl}


\bibitem[Zhang et~al\mbox{.}(2019)]%
        {zhang_actitouch_2019}
\bibfield{author}{\bibinfo{person}{Yang Zhang}, \bibinfo{person}{Wolf Kienzle}, \bibinfo{person}{Yanjun Ma}, \bibinfo{person}{Shiu~S. Ng}, \bibinfo{person}{Hrvoje Benko}, {and} \bibinfo{person}{Chris Harrison}.} \bibinfo{year}{2019}\natexlab{}.
\newblock \showarticletitle{{ActiTouch}: {Robust} {Touch} {Detection} for {On}-{Skin} {AR}/{VR} {Interfaces}}. In \bibinfo{booktitle}{\emph{Proceedings of the 32nd {Annual} {ACM} {Symposium} on {User} {Interface} {Software} and {Technology}}} \emph{(\bibinfo{series}{{UIST} '19})}. \bibinfo{publisher}{Association for Computing Machinery}, \bibinfo{address}{New York, NY, USA}, \bibinfo{pages}{1151--1159}.
\newblock
\showISBNx{978-1-4503-6816-2}
\urldef\tempurl%
\url{https://doi.org/10.1145/3332165.3347869}
\showDOI{\tempurl}


\bibitem[Zhang et~al\mbox{.}(2016)]%
        {zhang_skintrack_2016}
\bibfield{author}{\bibinfo{person}{Yang Zhang}, \bibinfo{person}{Junhan Zhou}, \bibinfo{person}{Gierad Laput}, {and} \bibinfo{person}{Chris Harrison}.} \bibinfo{year}{2016}\natexlab{}.
\newblock \showarticletitle{{SkinTrack}: {Using} the {Body} as an {Electrical} {Waveguide} for {Continuous} {Finger} {Tracking} on the {Skin}}. In \bibinfo{booktitle}{\emph{Proceedings of the 2016 {CHI} {Conference} on {Human} {Factors} in {Computing} {Systems}}}. \bibinfo{publisher}{ACM}, \bibinfo{address}{San Jose California USA}, \bibinfo{pages}{1491--1503}.
\newblock
\showISBNx{978-1-4503-3362-7}
\urldef\tempurl%
\url{https://doi.org/10.1145/2858036.2858082}
\showDOI{\tempurl}


\bibitem[Zhao et~al\mbox{.}(2025)]%
        {zhao_egopressure_2024}
\bibfield{author}{\bibinfo{person}{Yiming Zhao}, \bibinfo{person}{Taein Kwon}, \bibinfo{person}{Paul Streli}, \bibinfo{person}{Marc Pollefeys}, {and} \bibinfo{person}{Christian Holz}.} \bibinfo{year}{2025}\natexlab{}.
\newblock \showarticletitle{EgoPressure: A Dataset for Hand Pressure and Pose Estimation in Egocentric Vision}. In \bibinfo{booktitle}{\emph{Proceedings of the Computer Vision and Pattern Recognition Conference (CVPR)}}. \bibinfo{pages}{27727--27738}.
\newblock


\end{thebibliography}

\appendix

\end{document}